\documentclass[preprint,nofootinbib,floatfix,a4paper,prd,superscriptaddress]{revtex4-1}   	
\usepackage{geometry}                		
\usepackage[colorlinks,citecolor=blue]{hyperref}
\usepackage{amsmath}
\usepackage{mathrsfs}
\usepackage{bbm}
\usepackage{amsfonts}
\usepackage{amssymb}
\usepackage{latexsym}
\usepackage{graphicx}
\usepackage[english]{babel}
\usepackage{multirow}
\usepackage{float}
\usepackage{url}
\usepackage{slashed}
\usepackage{tabularx}
\usepackage{orcidlink}
\usepackage{appendix}

\newcommand{\be}{\begin{equation}}
\newcommand{\ee}{\end{equation}}
\newcommand{\ba}{\begin{array}}
\newcommand{\ea}{\end{array}}
\newcommand{\bea}{\begin{eqnarray}}
\newcommand{\eea}{\end{eqnarray}}

\usepackage[utf8]{inputenc}
\usepackage[T1]{fontenc}

\def  \bcen   {\begin{center}}
\def  \ecen   {\end{center}}
\def  \beq    {\begin{equation}}
\def  \eeq    {\end{equation}}

\def\la   {\lambda}

\def\lee { \left( }
\def\rii { \right) }

\def\to {\rightarrow}
\def\lphp {\la^\prime_{H\Phi}}

\newcommand{\TSUa}{\affiliation{\small Tsung-Dao Lee Institute \& School of Physics and Astronomy, Shanghai Jiao Tong University, Shanghai 200240, China }}

\newcommand{\PU}{\affiliation{\small Phenikaa Institute for Advanced Study, Phenikaa University, Yen Nghia, Ha Dong, Hanoi 100000, Vietnam}}

\newcommand{\AS}{\affiliation{\small Institute of Physics, Academia Sinica, Nangang, Taipei 11529, Taiwan}}

\begin{document}
\title{
\small When The Standard Model Higgs Meets Its Lighter 95 GeV Twin}

\author{ Abdesslam Arhrib}
\email{aarhrib@gmail.com}
\affiliation{\small Abdelmalek Essaadi University, Faculty of Sciences and Techniques, Tanger, Morocco }

\author{ Khiem Hong Phan }
\email{phanhongkhiem@duytan.edu.vn}
\affiliation{\small Institute of Fundamental and Applied Sciences, Duy Tan University, Ho Chi Minh City 70000,
Vietnam}
\affiliation{\small Faculty of Natural Sciences, Duy Tan University, Da Nang City 50000, Vietnam}

\author{\\ Van Que Tran 
\orcidlink{0000-0003-4643-4050}}
\email{vqtran@gate.sinica.edu.tw} \TSUa \PU \AS

\author{Tzu-Chiang Yuan \orcidlink{0000-0001-8546-5031} \, }
\email{tcyuan@phys.sinica.edu.tw} \AS

\begin{abstract}
Recent reports from the Large Hadron Collider indicate two excesses: one in the lighter Higgs mass region around 95 GeV, and another in the rare $Z \gamma$ final state of the Standard Model (SM) 125 GeV Higgs decay. These anomalies are analyzed within the minimal gauged two-Higgs-doublet model (G2HDM). A viable parameter space in G2HDM is identified that can account for both excesses. Within the viable parameter space, we find a strong correlation between the signal strengths of the SM 125 GeV Higgs decays into $\gamma \gamma$ and $Z \gamma$ modes. However, this correlation does not extend to the lighter 95 GeV Higgs.

\end{abstract}

\maketitle

\section{Introduction}
\label{secIntro}

Facing the mounting exabyte of physics data collected by the Large Hadron Collider (LHC) since its first collisions in 2010, the Standard Model (SM) is holding up remarkably well, successfully withstanding all challenges posed by our experimental colleagues.

Although the LHC has reported occasional middling 2$\sim$3$\sigma$ excesses over the years, sparking excitement in the community, these moments of joy have been short-lived as the excesses quickly faded with increased statistical data. Nonetheless, few doubt that there must be new physics beyond the Standard Model (BSM). From a bottom-up perspective, the experiments on neutrino oscillations suggest the need for right-handed neutrinos. Dark matter and dark energy, absent in the SM, are crucial for the evolution and history of our observable universe. From a top-down perspective, compelling arguments for new physics arise from grand unification, gauge hierarchy problem, and theories of everything such as string theory.

Recently, CMS Collaboration has reported an excess in the light Higgs-boson search in the di-photon ($\gamma\gamma$) decay
mode at about 95.4 GeV based on the 8 TeV data and the full Run 2 data set at 13 TeV with the local significance of $2.9 \sigma$ \cite{CMS-PAS-HIG-20-002}.  
The corresponding signal strength is given as 
\begin{equation}
\mu_{\gamma \gamma}^{\mathrm{CMS}}=\frac{\sigma^{\exp }(g g \rightarrow h \rightarrow \gamma \gamma)}{\sigma^{\mathrm{SM}}(g g \rightarrow H \rightarrow \gamma \gamma)}=0.33_{-0.12}^{+0.19}  \; ,
\end{equation}
where $\sigma^{\rm exp}$ denotes the cross section for a hypothetical scalar boson $h$ with the mass is 95.4 GeV, 
and $H$ is the SM Higgs.

ATLAS Collaboration recently also presented the result of the search for new neutral scalars in the di-photon final state with mass window from 66 GeV to 110 GeV, using full Run 2 data collected at 13 TeV \cite{ATLAS:seminar, ATLAS-CONF-2023-035}. 
ATLAS observed an excess at the same mass value as reported by the CMS with the local significance is $1.7 \sigma$. The corresponding signal strength is given as \cite{Biekotter:2023oen}
\begin{equation}
\mu_{\gamma \gamma}^{\mathrm{ATLAS}}=\frac{\sigma^{\exp }(g g \rightarrow h \rightarrow \gamma \gamma)}{\sigma^{\mathrm{SM}}(g g \rightarrow H \rightarrow \gamma \gamma)}=0.18_{-0.1}^{+0.1} \; .
\end{equation}

The combined local significance from ATLAS and CMS is $3.1 \sigma$ and the signal strength is~\cite{Biekotter:2023oen}
\begin{equation}
\label{mugammagammaCombined}
\mu_{\gamma \gamma}^{\exp }=\mu_{\gamma \gamma}^{\text {ATLAS }+\mathrm{CMS}}= 0.24_{-0.08}^{+0.09} \; .
\end{equation}

Using the full Run 2 data set, CMS reported another local
excess with a significance of $3.1 \sigma$ ($2.6 \sigma$) for light Higgs with mass $\sim 100$ GeV ($95$ GeV) produces from gluon-gluon fusion and subsequently decays to di-tau ($\tau^+\tau^-$) final state~\cite{CMS-PAS-HIG-21-001}. 
The signal strength for the scalar mass at $95$ GeV is given as
\begin{equation}
\label{mutautauCMS}
\mu_{\tau \tau}^{\mathrm{exp}}=\frac{\sigma^{\exp }(g g \rightarrow h \rightarrow \tau^{+} \tau^{-})}{\sigma^{\mathrm{SM}}(g g \rightarrow H \rightarrow \tau^{+} \tau^{-})} = 1.2 \pm {0.5}  \; . 
\end{equation}
Note that ATLAS has not yet reported a search in the di-tau final state that covers the mass range around 95 GeV.

In addition, searches for a low-mass scalar boson were previously carried out at LEP. A local significance excess of $2.3 \sigma$ for the light scalar mass of about 98 GeV in the process $e^+ e^- \to Z(H \to b \overline{b})$ and the corresponding signal strength of~\cite{LEPWorkingGroupforHiggsbosonsearches:2003ing}
\footnote{The signal strength may be altered if the scalar mass shifts to approximately $95$ GeV~\cite{Janot:2024ryq}. However, in this analysis, we assume that the signal strength remains unchanged. } 
\begin{equation}
\label{mubb}
    \mu^{\rm exp}_{b b} = 0.117 \pm 0.057 \; .       
\end{equation}

At face values, the signal strengths in (\ref{mugammagammaCombined}), (\ref{mutautauCMS}) and
(\ref{mubb}) indicate that the 95 GeV Higgs, if its existence is confirmed, would be very much SM-like for the di-tau mode but rather non-SM like for both the di-photon and $ b \bar b$ modes.
Excesses in these channels are strong motivation for BSM. Indeed in light of these excesses, many BSMs have been studied in recent years, {\it e.g.}~\cite{Sachdeva:2019hvk,Biekotter:2023jld,Biekotter:2022jyr,Iguro:2022dok,Iguro:2022fel,Coloretti:2023wng,Azevedo:2023zkg,Escribano:2023hxj,Biekotter:2023oen,Belyaev:2023xnv,Ashanujjaman:2023etj,Bhattacharya:2023lmu,Aguilar-Saavedra:2023tql,Dutta:2023cig,Banik:2023vxa,Ellwanger:2023zjc,Cao:2023gkc,Borah:2023hqw,Ahriche:2023hho,Arcadi:2023smv,Ahriche:2023wkj,Chen:2023bqr,Coloretti:2023yyq,Li:2023kbf,Dev:2023kzu,He:2024sma,Benbrik:2024ptw}.

More recently, both ATLAS and CMS Collaborations~\cite{ATLAS:2023yqk} reported an analysis for the first evidence of the rare decay mode $H \to Z\gamma$, where the $Z$ boson decays into a $e^+e^-$ or $\mu^+\mu^-$ pair. 
The number of events is found twice as many as predicted by the SM. To be more precise, the combined observed signal yield is
\begin{equation}
\label{muZgamma}
    \mu_{Z\gamma}^{\rm exp} = \frac{ \sigma^{\rm exp} (g g \to H \to Z \gamma) }{ \sigma^{\rm SM} (g g \to H \to Z \gamma) }
    = 2.2 \pm 0.7 \; ,
\end{equation}
with a 3.4$\sigma$ statistical significance.
Currently, the data are insufficient to rule out the possibility that the discrepancies are merely due to random variation. Nonetheless, they present an opportunity for new physics enthusiasts to explore potential new phenomena.

In general, explaining the excesses observed in (\ref{mugammagammaCombined}), (\ref{mutautauCMS}), and (\ref{mubb}) requires the presence of a new scalar boson with a mass around 95 GeV and an enhanced signal strength, particularly in the di-tau final state. A key requirement is that this scalar must have a sufficiently large production cross-section at the LHC (for the di-photon and di-tau excesses) and at LEP (for the \( b\bar{b} \) excess), along with significant decay rates into the corresponding SM particles. This demands a sizable coupling of the new scalar to SM particles and/or additional loop-induced contributions that enhance its production and decay processes.
Meanwhile, addressing the excess in (\ref{muZgamma}) requires a significant enhancement in the signal strength of the 125 GeV Higgs boson decaying into \( Z\gamma \) compared to its SM expectation, while remaining consistent with current LHC Higgs data.

In this paper, we study the excesses (\ref{mugammagammaCombined}), (\ref{mutautauCMS}), (\ref{mubb}) and (\ref{muZgamma}) in the framework of minimal gauged two-Higgs-doublet model (G2HDM) which contains a predominantly $SU(2)_L$ singlet scalar $h_1$, a mixture of a hidden scalar with the SM Higgs boson, can become the 95 GeV Higgs boson candidate. The orthogonal combination $h_2$ will be identified as the observed 125 GeV Higgs boson.
Besides the SM $W^\pm$ boson and top quark $t$ contributions, the presence of charged heavy hidden fermions $(f^H)$ and charged Higgs $H^\pm$ can provide additional contributions to the production and decay through one-loop diagrams for both Higgs bosons. Thus it is natural that the signal strengths for these two Higgs bosons deviate from their SM expectations.

This paper is organized as follows. We will give a brief review of the minimal G2HDM in Section~\ref{sec:MG2HDM}, followed by a discussion in Section~\ref{sec:SignalStrengths} for the computation of signal strengths in the model. Numerical studies of scanning the parameter space in G2HDM are presented in Section~\ref{sec:Numbers}. We conclude in Section~\ref{sec:Conclusions}.
For convenience, Appendix~\ref{appA} collects the detailed decay rates for $\gamma\gamma$, $Z\gamma$ and $gg$ modes of the Higgs bosons. We also discuss our numerical study of the two loop functions entered in the Higgs decay amplitudes.

\section{The Minimal G2HDM and its constraint}
\subsection{The model}
\label{sec:MG2HDM}

The crucial idea of the original G2HDM~\cite{Huang:2015wts} was to embed the two Higgs doublets $H_1$ and $H_2$ in the popular scalar dark matter model, the inert two-Higgs-doublet model (I2HDM), into a 2-dimensional irreducible representation of a hidden gauge group $SU(2)_H \times U(1)_X$, a dark replica of the SM electroweak group $SU(2)_L \times U(1)_Y$. 
Besides the two Higgs doublets in I2HDM, the original G2HDM also introduced two hidden scalar multiplets, one doublet ($\Phi_H$) and one triplet ($\Delta_H$), to generate the hidden particle mass spectra.
The hidden gauge group acts horizontally on the two Higgs doublets in I2HDM.

Various refinements~\cite{Arhrib:2018sbz,Huang:2019obt,Chen:2019pnt} and 
collider phenomenology~\cite{Huang:2015rkj,Chen:2018wjl,Huang:2017bto} were pursued subsequently with the same particle content as in 
the original model. 
Recently~\cite{Ramos:2021omo,Ramos:2021txu}, it has been realized that removing the hidden triplet scalar field $\Delta_H$ in the model without jeopardizing a realistic hidden particle mass spectra. 
Interpretation~\cite{Tran:2022yrh}
~\footnote{Recently high-precision measurement of the $W$ boson mass with the CMS experiment at the LHC running at 13 TeV has reported~\cite{CMS:2024lrd} $m_W = 80360.2 \pm 9.9$ MeV, in agreement with the SM prediction.}
of the $W$ boson mass measurement at the CDF II~\cite{CDF:2022hxs}, FCNC processes $l_i \to l_j \gamma$~\cite{Tran:2022cwh}, $b \to s \gamma$~\cite{Liu:2024nkl}, and on the interplay between gravitational wave and dark matter signals via a two-step first order phase transition~\cite{Ramsey-Musolf:2024zex} have been recently studied within the framework of this minimal G2HDM. 
Thus, the model stands out in its ability to address a wide range of fundamental physics topics, encompassing collider physics, dark matter, and the broader dark sector, while also providing rich phenomenology in flavor physics and gravitational wave signals. In this work, we will specifically concentrate on the minimal G2HDM in the context of collider physics, exploring its distinct signatures and potential experimental implications, motivated by the anomalies discussed in the previous section.

The gauge group of G2HDM is 
$$
\mathcal G = SU(3)_C \times SU(2)_L \times SU(2)_H \times U(1)_Y \times U(1)_X \; .
$$
The minimal particle representations under $\mathcal G$ are as follows:

\noindent
\underline{Spin 0 Bosons}:
$$
\mathcal H = \left( H_1 \;\; H_2 \right)^{\rm T} \sim  \left( {\bf 1}, {\bf 2}, {\bf 2}, \frac{1}{2}, { \frac{1}{2} }\right) \; , \; $$
$$\Phi_H \sim \left( {\bf 1}, {\bf 1}, {\bf 2}, 0, { \frac{1}{2} } \right) \; ; 
$$

\noindent
\underline{Spin 1/2 Fermions}:

\underline{Quarks}
$$Q_L=\left( u_L \;\; d_L \right)^{\rm T} \sim \left(  {\bf 3}, {\bf 2}, {\bf 1}, \frac{1}{6}, 0 \right) \; , \; $$
$$U_R=\left( u_R \;\; u^H_R \right)^{\rm T} \sim  \left( {\bf 3}, {\bf 1}, {\bf 2}, \frac{2}{3},  \frac{1}{2}  \right) \; , $$
$$D_R=\left( d^H_R \;\; d_R \right)^{\rm T} \sim \left( {\bf 3}, {\bf 1}, {\bf 2},  -\frac{1}{3}, - \frac{1}{2}  \right) \; ; $$
$$u_L^H \sim \left(  {\bf 3}, {\bf 1}, {\bf 1},  \frac{2}{3}, 0 \right) \; , \; d_L^H \sim \left(  {\bf 3}, {\bf 1}, {\bf 1}, -\frac{1}{3}, 0 \right) \; ; $$

\underline{Leptons}
$$L_L=\left( \nu_L \;\; e_L \right)^{\rm T} \sim \left( {\bf 1}, {\bf 2}, {\bf 1},  -\frac{1}{2}, 0 \right) \; , \; $$
$$N_R=\left( \nu_R \;\; \nu^H_R \right)^{\rm T} \sim \left( {\bf 1}, {\bf 1}, {\bf 2},  0,  \frac{1}{2}  \right)  \; , $$
$$E_R=\left( e^H_R \;\; e_R \right)^{\rm T} \sim \left( {\bf 1}, {\bf 1}, {\bf 2},  -1, - \frac{1}{2}  \right) \; ; $$
$$\nu_L^H \sim \left( {\bf 1}, {\bf 1}, {\bf 1},  0, 0 \right) \; , \; e_L^H \sim \left( {\bf 1}, {\bf 1}, {\bf 1},  -1, 0 \right) \; .$$
We assume three families of matter fermions in minimal G2HDM and the family indices are often omitted. In addition to the SM gauge fields $W_i$ ($i=1,2,3$) of $SU(2)_L$ and $B$ of $U(1)_Y$, the hidden gauge fields of $SU(2)_H$ and $U(1)_X$ are denoted as $W^\prime_i$ ($i=1,2,3)$ 
and $X$ respectively.

One of the nice features of G2HDM is the presence of the accidental $h$-parity~\cite{Chen:2019pnt} such that all the SM particles can be assigned to be even. The $h$-parity odd particles in G2HDM are $W^{\prime \, (p,m)}=(W^\prime_1 \mp i W^\prime_2)/\sqrt{2}$, $D^{(*)}$ and ${\tilde G}^{(*)}$ (the two orthogonal combinations of the complex neutral component in $H_2$ and the hidden complex Goldstone field in $\Phi_H$), $H^\pm$, and all new heavy fermions collectively denoted as $f^H$.
Among them, $W^{\prime \, (p,m)}$, $D^{(*)}$, and $\nu^H$ are electrically neutral and hence any one of them can be a dark matter (DM) candidate.
Phenomenology of a complex scalar $D^{(*)}$ as DM was studied in details in~\cite{Chen:2019pnt,Dirgantara:2020lqy} 
and for low mass $W^{\prime \, (p,m)}$ as DM, see~\cite{Ramos:2021omo,Ramos:2021txu}. A pure gauge-Higgs sector with $W^{\prime \, (p,m)}$ as self-interacting dark matter was also studied recently in~\cite{Tran:2023lzv}.
For further details of G2HDM, we refer our readers to the earlier works~\cite{Huang:2019obt,Arhrib:2018sbz,Ramos:2021omo,Ramos:2021txu}. 
Phenomenology of a new heavy neutrino $\nu^H$ as DM in the model, which is necessarily  implying
both DM and neutrino physics, has yet to be explored.

The most general renormalizable Higgs potential invariant under both the SM $SU(2)_L\times U(1)_Y$ and the hidden $SU(2)_H \times  U(1)_X$  
is given by
\begin{align}\label{eq:V}
V = {}& - \mu^2_H   \left({\mathcal H}^{\alpha i}  {\mathcal H}_{\alpha i} \right)
+  \lambda_H \left({\mathcal H}^{\alpha i}  {\mathcal H}_{\alpha i} \right)^2  
+ \frac{1}{2} \lambda'_H \epsilon_{\alpha \beta} \epsilon^{\gamma \delta}
\left({\mathcal H}^{ \alpha i}  {\mathcal H}_{\gamma  i} \right)  \left({\mathcal H}^{ \beta j}  {\mathcal H}_{\delta j} \right)  \nonumber \\
{}&- \mu^2_{\Phi}   \Phi_H^\dag \Phi_H  + \la_\Phi \lee \Phi_H^\dag \Phi_H  \rii^2  \\
{}&
+\lambda_{H\Phi} \lee {\mathcal H}^\dag {\mathcal H}  \rii  \lee \Phi_H^\dag \Phi_H \rii  
 + \lambda^\prime_{H\Phi} \lee {\mathcal H}^\dag \Phi_H  \rii  \lee \Phi_H^\dag {\mathcal H} \rii, \nonumber
\end{align}
where  ($\alpha$, $\beta$, $\gamma$, $\delta$) and ($i$, $j$) refer to the $SU(2)_H$ and $SU(2)_L$ indices respectively, 
all of which run from 1 to 2, and ${\mathcal H}^{\alpha i} = {\mathcal H}^*_{\alpha i}$. We note that every term in the above potential $V$ is self-Hermitian. Therefore all parameters in (\ref{eq:V}) are real and no CP violation can be arise from the scalar potential. This tree-level potential augmented by its high temperature corrections has been analyzed recently in \cite{Ramsey-Musolf:2024zex} to realize a 2-step strong first-order electroweak phase transition.

To achieve spontaneous symmetry breaking (SSB) in the model, we follow standard lore to parameterize the Higgs fields in the doublets linearly as
\begin{widetext}
\begin{eqnarray}
\label{eq:scalarfields}
H_1 = 
\begin{pmatrix}
G^+ \\ \frac{ v + h_{\rm SM}}{\sqrt 2} + i \frac{G^0}{\sqrt 2}
\end{pmatrix}
, \;
H_2 = 
\begin{pmatrix}
 H^+  \\  H_2^0 
\end{pmatrix}
, \;
\Phi_H = 
\begin{pmatrix}
G_H^p  \\ \frac{ v_\Phi + \phi_H}{\sqrt 2} + i \frac{G_H^0}{\sqrt 2}
\end{pmatrix}
\; \; \;
\end{eqnarray}
\end{widetext}
where $v$ and $v_\Phi$ are the only non-vanishing vacuum expectation values (VEVs)
in the $H_1$ and $\Phi_{H}$ doublets respectively. $v = 246$ GeV is the SM VEV, and $v_\Phi$ is an unknown hidden VEV. We note that $h$-parity would be broken spontaneously should $\langle H_2 
\rangle$ develops a VEV. This is undesirable since we want a DM candidate to address DM physics at low energy.

In G2HDM, the SM charged gauge boson $W^\pm$ does not mix with $W^{\prime \, (p,m)}$ and its mass is the same as in SM: $m_W = g v/2$.
However the SM neutral gauge boson $Z_{\rm SM}$ will in general mix further with the gauge field $W^{\prime}_3$ associated with the third generator of $SU(2)_H$ 
and the $U(1)_X$ gauge field $X$ via the following mass matrix
\be
\mathcal M_Z^2 =  
\begin{pmatrix}
m^2_{Z} & - \frac{1}{2} g_H v m_{Z} & - { \frac{1}{2} } g_X v m_{Z} \\
 - \frac{1}{2} g_H v m_{Z} & m^2_{W^\prime} & { \frac{1}{4} } g_H g_X v_{-}^2 \\
- {\frac{1}{2} } g_X v m_{Z} &   { \frac{1 }{4} } g_H g_X v_{-}^2 & {\frac{1}{4} } g_X^2 v_{+}^2 + M_X^2
\end{pmatrix} \; ,
\label{MsqZs}
\ee
where 
\bea
\label{vpm}
v_{\pm}^2 & = & \left( v^2 \pm v^2_\Phi \right) \; , \\
 \label{mzsm}
 m_{Z} & = &  \frac{1 }{2}  v \sqrt{ g^2 + g^{\prime \, 2} }\; , \\
 \label{mwprime}
 m_{W^\prime} & = & \frac{1}{2} g_H \sqrt{ v^2 + v_\Phi^2 } \; ,
\eea
and $M_X$ is the Stueckelberg mass for the $U(1)_X$. 

The real and symmetric mass matrix $\mathcal M_Z^2 $ in (\ref{MsqZs}) can be diagonalized by a 3 by 3 orthogonal matrix ${\mathcal O}^G$,
{\it i.e.} $(\mathcal O^{G})^{\rm T} \mathcal M_Z^2 \mathcal O^G = {\rm Diag}(m^2_{Z_1}, m^2_{Z_2},m^2_{Z_3})$, where $m_{Z_i}$ is 
the mass of the physical fields $Z_i$ for $ i=1,2,3$.
We will identify $Z_1 \equiv Z$ to be the neutral gauge boson resonance with a mass  of  91.1876  GeV observed at LEP~~\cite{Zyla:2020zbs}. 
The lighter/heavier of the other two states is the dark photon ($\gamma^\prime$)/dark $Z$ ($Z^\prime$). These neutral gauge bosons are $h$-parity even in the model,
despite the adjective `dark' are used for the other two states. 
We note that these neutral gauge bosons can decay into SM particles and thus they can be constrained by experimental data, including the electroweak precision measurement at the $Z$ pole physics from LEP, searches for dark $Z$ and dark photon at colliders, beam-dump experiments, and astrophysical observations.
The DM candidate considered in this work is $W^{\prime \, (p,m)}$, which is electrically neutral but carries one unit of dark charge and
chosen to be the lightest $h$-parity odd particle in the parameter space.

In G2HDM there are mixings effects of the two doublets $H_1$ and $H_2$ with the hidden doublet $\Phi_H$.
The neutral components $h_{\rm SM}$ and $\phi_H$ in $H_1$ and $\Phi_H$ respectively are both $h$-parity even.
They mix to form two physical Higgs fields $h_1$ and $h_2$
\be
\label{eq:mixingmatrix}
\left(
\begin{matrix}
h_{\rm SM} \\
\phi_H
\end{matrix}
\right)
= 
{\mathcal O}^S \cdot 
\left(
\begin{matrix}
h_1 \\
h_2
\end{matrix}
\right)
=
\left( 
\begin{matrix}
 \cos \theta_1  &  \sin \theta_1 \\
- \sin \theta_1  &  \cos \theta_1 
\end{matrix}
\right) \cdot 
\left(
\begin{matrix}
h_1 \\
h_2
\end{matrix}
\right)
\; .
\ee
The mixing angle $\theta_1$ is given by
\be
\label{theta1}
\tan 2 \theta_1 = \frac{\lambda_{H\Phi} v v_\Phi }{ \lambda_\Phi v^2_\Phi - \lambda_H v^2 } \; .
\ee
The masses of $h_1$ and $h_2$ are given by
\bea
\label{massesh1h2}
m_{h_1,h_2}^2 &=& \lambda_H v^2 + \lambda_\Phi v_\Phi^2 \nonumber \\
& \mp &\sqrt{\lambda_H^2 v^4 + \lambda_\Phi^2 v_\Phi^4 + \left( \lambda^2_{H\Phi}  - 2 \lambda_H \lambda_\Phi \right) v^2 v_\Phi^2 }\,.  \hspace{0.6cm}
\eea
Depending on its mass, either $h_1$ or $h_2$ is designated as the observed Higgs boson at the LHC. Currently, the most precise measurement of the Higgs boson mass is $125.38 \pm 0.14$ GeV~\cite{CMS:2020xrn}. In this analysis, $h_1$ and $h_2$ are identified as the lighter and SM Higgs bosons with masses approximately $95$ GeV and $125$ GeV, respectively, to address the LHC excesses. For clarity in denoting the masses of the scalars, we use the notation $h_{95} \equiv h_{1}$ and $h_{125} \equiv h_{2}$ in the following.

The complex fields $H_2^{0 \, *}$ and $G^p_H$ in $H_2$ and $\Phi_H$ respectively are both $h$-parity odd. They
mix to form a physical dark Higgs $D^*$ and an unphysical Goldstone field $\tilde G^*$ absorbed by the $W^{\prime \, p}$
\be
\label{H20field}
\begin{pmatrix}
G^m_H \\
H_2^{0} 
\end{pmatrix}
= \mathcal O^D \cdot
\begin{pmatrix}
 \tilde G \\
 D 
 \end{pmatrix}
=
\begin{pmatrix}
 \cos \theta_2 & \sin \theta_2  \\
- \sin \theta_2  & \cos \theta_2 
\end{pmatrix} 
\cdot
\begin{pmatrix}
 \tilde G \\
 D 
 \end{pmatrix}
 \; .
\ee
The mixing angle $\theta_2$ satisfies
\be
\label{theta2}
\tan 2 \theta_2 = \frac{2 v v_\Phi}{v^2_\Phi - v^2} \; ,
\ee
and the mass of $D$ is 
\be
m_D^2 = \frac{1}{2} \lambda^\prime_{H\Phi} \left( v^2 + v_\Phi^2 \right) \; .
\ee
In the Feynman-'t Hooft gauge the Goldstone field $\tilde G^*$ $(\tilde G )$ has the same mass as the $W^{\prime \, p}$ ($W^{\prime \, m}$) which is given by 
(\ref{mwprime}).
Finally the charged Higgs $H^\pm$ is also $h$-parity odd and has a mass
\be
m^2_{H^\pm} = \frac{1}{2} \left( \lambda^\prime_{H\Phi} v^2_\Phi - \lambda^\prime_H v^2 \right) \; .
\ee

One can do the inversion to express the fundamental parameters in the scalar potential in terms of the particle masses and mixing angle $\theta_1$~\cite{Ramos:2021txu,Ramos:2021omo}: 
\bea
\lambda_H & = & \frac{1}{2 v^2} \left( m^2_{h_{95}} \cos^2 \theta_1 + m^2_{h_{125}} \sin^2 \theta_1 \right) \; , \\
\lambda_\Phi & = & \frac{1}{2 v_\Phi^2} \left( m^2_{h_{95}} \sin^2 \theta_1 + m^2_{h_{125}} \cos^2 \theta_1 \right) \; , \\
\lambda_{H\Phi} & = & \frac{1}{2 v v_\Phi} \left( m^2_{h_{125}} - m^2_{h_{95}} \right) \sin \left( 2 \theta_1 \right) \; , \\
\lambda^\prime_{H\Phi} & = & \frac{ 2  m_D^2 }{v^2 + v_\Phi^2} \; , \\
\lambda^\prime_{H} & = & \frac{2}{v^2} \left( \frac{m_D^2 v_\Phi^2 }{ v^2 + v_\Phi^2 } - m^2_{H^\pm}\right) \; .
\eea
From (\ref{mwprime}), we also have
\be
g_H  =   \frac{ 2 m_{W'} } { \sqrt{v^2 + v^2_{\Phi} }}  \;. 
\ee
Thus one can elegantly use the masses $m_{h_{95}}$, $m_{W^\prime}$, $m_D$ and $m_{H^\pm}$, mixing angle $\theta_1$ and VEV $v_\Phi$ as input in our numerical scan. 

The connector sector linking the SM particles and the DM $W^{\prime (p,m)}$ consists of the $h$-parity even or odd particles. Specifically, we have 
$\gamma$, $Z_i (i=1,2,3)$, $h_{125}$ and $h_{95}$ in the $s$-channel, and $D$, $H^\pm$, new heavy fermions $f^H$s 
and DM $W^{\prime (p,m)}$ itself 
in the $t$-channel and/or $u$-channel.

\subsection{The constraints}
\label{sec:constraints}
The analysis of constraints on the model have been carried out previously in~\cite{Ramos:2021txu,Ramos:2021omo,Tran:2022cwh}, which include theoretical constraints on the scalar potential, experimental constraints from the Higgs measurements at the LHC, electroweak precision data, dark photon and dark matter searches.
In the following, we will summarize these constraints: 

{\underline {\it Theoretical constraints: }}

(a) Vacuum Stability: 
To guarantee that the scalar potential has a minimum value, 
we use the copositivity conditions suggested by~\cite{Arhrib:2018sbz}, which provide the following set of constraints 
on the scalar potential parameters 
\be
\widetilde \lambda_H (\eta) \geq 0, \;\; \lambda_\Phi \geq 0 \;\; {\rm and} \;\;
\widetilde \lambda_{H\Phi}(\xi) + 2 \sqrt{\widetilde \lambda_H (\eta) \lambda_\Phi}  \geq  0,
\ee
where $\widetilde \lambda_H (\eta) \equiv \lambda_H + \eta \lambda^\prime_H$ and
${\widetilde  \lambda_{H\Phi}}(\xi) \equiv \lambda_{H\Phi} + \xi \lphp $ with $0 \leq \xi \leq 1$ and $-1 \leq \eta \leq 0$.

(b) Perturbative Unitarity: 
Another theoretical constraint from perturbative unitarity requires ~\cite{Arhrib:2018sbz, Ramos:2021txu, Ramos:2021omo}, 
\begin{align}
&\vert \lambda_H \vert, \vert  \lambda_\Phi \vert  \leq  4 \pi \;,
 \vert \lambda_{H\Phi} \vert  \leq  8 \pi \;,
 \vert \lambda^\prime_{H\Phi} \vert \;, \vert \lambda_H^\prime \vert  \leq  8 \sqrt 2 \pi \; , \\
& \vert 2 \lambda_H \pm \lambda^\prime_H \vert \leq 8 \pi \;,
\vert \lambda_{H\Phi} + \lambda_{H\Phi}^\prime \vert \leq  8 \pi \; ,\\
& \left| ( \lambda_H + \lambda_H^\prime/2 +
\lambda_\Phi) \pm \sqrt{2\lambda^{\prime 2}_{H\Phi} +
( \lambda_H + \lambda_H^\prime/2 - \lambda_\Phi)^2} \right|  \leq  8 \pi \;,  \\
& \left| (5 \lambda_H - \lambda_H^\prime/2 + 3 \lambda_\Phi) 
\pm \sqrt{(5 \lambda_H - \lambda_H^\prime/2 - 3\lambda_\Phi)^2 
+ 2 (2 \lambda_{H\Phi} + {\lambda}'_{H\Phi} )^2} \right|  \leq  8 \pi \; .
\end{align}

{\underline{\it Higgs data constraints }}:
Due to the mixing between the $h$-parity even scalars shown in (\ref{eq:mixingmatrix}), the couplings of the $125$ GeV Higgs boson to SM particles are modified as compared with those in the SM.    
Moreover, with the presence of hidden charged fermions $f^{\rm H}$ and charged Higgs boson $H^\pm$ in the model, the one-loop process $h \to \gamma \gamma$ can be also altered~\cite{Tran:2022cwh}. Here, we employ the Higgs signal strength data from CMS~\cite{CMS-PAS-HIG-19-005} experiments to constraint the model parameters. 

In additional, if it is kinematically allowed, the Higgs boson can decay invisibly into a pair of dark matter $W^{\prime (p,m)}$. The invisible branching ratio can be given by \cite{Ramos:2021txu,Ramos:2021omo}
\be
{\rm BR} ({h_{125} \to {\rm inv}}) =\frac{ \Gamma({h_{125} \to W^{\prime p} W^{\prime m}})}{\Gamma_{h_{125}}} \; ,
\ee 
where 
$\Gamma_{h_{125}} \simeq \cos^2{\theta_1} \Gamma^{\rm SM}_{h} + \Gamma({h_{125} \to W^{\prime p} W^{\prime m}})$ is the total decay width of the Higgs boson, $\Gamma^{\rm SM}_{h} = 4.1$ MeV is the SM decay width 
and $\Gamma({h_{125} \to W^{\prime p} W^{\prime m}})$ is its invisible decay width. The latter is given by
\be
\Gamma ({h_{125} \to W^{\prime p} W^{\prime m}}) = \frac{g_H^4 \left( v \cos \theta_1 - v_\Phi \sin \theta_1 \right)^2 }{256 \pi} 
\frac{m_{h_{125}}^3}{m_{W'}^4} \left( 1- \tau_{W'} +\frac{3}{4}  \tau_{W'}^2\right) 
\sqrt{1- \tau_{W'} } \; ,
\label{eq:invdecaywidth}
\ee
with $\tau_{W'} = 4 m_{W'}^2 / m_{h_{125}}^2$. 
We used the recent limit of ${\rm BR}({h_{125} \to{ \rm inv}}) < 0.145$ at $95 \%$ C.L. from ATLAS experiment~\cite{ATLAS:2022yvh} where the Higgs boson production cross-section via vector boson fusion was assumed to be comparable to the SM prediction.

\underline{{\it Electroweak precision data and dark photon/$Z^\prime$ constraints: }}

We take into account constraints from electroweak precision data at the $Z$ pole~\cite{Zyla:2020zbs}, as well as from  $Z'$~\cite{ATLAS:2019erb} and dark photon physics~\cite{Fabbrichesi:2020wbt}. 
For the dark photon and $Z'$ masses $m_{A', Z'} < m_Z$, 
these constraints require the new gauge couplings $g_H$ and $g_X$ to be less than $\sim 10^{-3}$~\cite{Ramos:2021txu,Ramos:2021omo}. 

We also take into account the constraints from $W$ boson mass measurements. The mass of $W$ boson in the model can be obtained in terms of the oblique parameters $\cal S$, $\cal T$, and $\cal U$~\cite{Peskin:1991sw} as
\begin{equation}
m^2_W - m^2_{W,{\rm SM}} = \frac{\alpha_{\text{EM}}c^2_W m^2_Z}{c^2_W - s^2_W}\left[ -\frac{\cal S}{2} + c^2_W {\cal T} + \frac{c^2_W - s^2_W}{4s^2_W}{\cal U}\right],
\end{equation}
where $c_W$ ($s_W$) is the cosine (sine) of weak mixing angle, $\alpha_{\text{EM}}$ is the fine structure constant, $m_Z$ is the $Z$ boson mass and $m_{W,{\rm SM}}$ is the $W$ boson mass in the SM. The oblique parameters $\cal S$, $\cal T$, and $\cal U$ in the model can be found in Ref.~\cite{Tran:2022yrh}. We use the latest updated overall average result for $W$ boson mass (without the measurement of the CDF from Run II at the Tevatron~\cite{CDF:2022hxs}) from the Particle Data Group \cite{ParticleDataGroup:2024cfk}, 
\be
m_W = 80.366 \pm 0.012 \; {\rm GeV}. 
\ee
Recently~\footnote{The CMS Collaboration, \href
{https://cds.cern.ch/record/2910372/files/SMP-23-002-pas.pdf}{CMS PAS SMP-23-002}.}, a very precise $W$ boson mass measurement of $80360.2 \pm 9.9 \; {\rm MeV}$
is in agreement with the SM prediction.

\underline{\it DM constraints: }
The DM constraints have been considered in our previous studies \cite{Ramos:2021txu,Ramos:2021omo}. In particular, we required a correct DM relic density $\Omega_{\rm DM} h^2 = 0.120 \pm 0.001$ from the Planck Collaboration~\cite{Aghanim:2018eyx}. Moreover, we used the recent upper limits from DM direct detection experiments including
 CRESST III \cite{Angloher:2017sxg}, DarkSide-50 \cite{Agnes:2018ves,DarkSide-50:2022qzh}, XENON1T \cite{XENON:2018voc, Aprile:2019xxb}, {XENONnT \cite{XENON:2023cxc} }, PandaX-4T \cite{PandaX-4T:2021bab} and LZ \cite{LZ:2022lsv}. 

All the constraints summarized above are included in this current work.

\section{The signal strengths}\label{sec:SignalStrengths}

In this section we show the signal strengths of the lighter scalar boson $h_{95}$ decays into di-photon and $\tau^+ \tau^-$ from gluon-gluon fusion, and into $b\bar{b}$ from Higgs-strahlung process. 
The signal strengths from the gluon-gluon fusion process can be given as 
\bea
\mu_{\gamma \gamma/\tau\tau } &=& \frac{\sigma(g g \to h_{95}) \times {\rm BR}(h_{95}\to \gamma \gamma/\tau^+ \tau^-)}{\sigma^{\rm SM}(g g \to H) \times  {\rm BR}^{\rm SM}(H\to \gamma \gamma/\tau^+ \tau^-)|{_{m_{H} = m_{h_{95}}}} } \, , \\
&=& \left. \frac{\Gamma^{\rm SM}_H}{\Gamma_{h_{95}}} \cdot \frac{\Gamma_{h_{95} \to gg} }{\Gamma^{\rm SM}_{H \to gg} } \cdot \frac{\Gamma_{h_{95} \to \gamma \gamma /\tau^+ \tau^-} }{\Gamma^{\rm SM}_{H \to \gamma \gamma /\tau^+ \tau^-}  } \right\vert_{m_H = m_{h_{95}}} \, ,
\eea
where $\Gamma^{\rm SM}$ is the SM total decay with of a hypothetical scalar with mass $m_H = m_{h_{95}} \simeq 95$ GeV, while $\Gamma_{h_{95}}$ is total decay width of $h_{95}$ boson. We follow Ref.~\cite{Choi:2021nql} to include radiative corrections for the decay widths $\Gamma^{\rm SM}$ and $\Gamma_{h_{95}}$.
Moreover, we have the decay width ratio of $\Gamma_{h_{95} \to \tau^+ \tau^-}/\Gamma^{\rm SM}_{H \to \tau^+ \tau^-} = \cos^2 \theta_1$. 
The decay width of $\Gamma_{h_{95} \to gg}$ and $\Gamma_{h_{95} \to \gamma \gamma}$ are given in the Appendix~\ref{appA}. 

The signal strength of the Higgs-strahlung process ($e^+ e^- \to Z h_{95} (h_{95} \to b \overline{b})$) is given by 
\be
\label{signalh1bb}
\mu_{bb} = \frac{\sigma (e^+ e^- \to Z h_{95})}{\sigma^{\rm SM} (e^+ e^- \to Z H)\vert_{ {m_{H} = m_{h_{95}}}} } \times \frac{{\rm BR} (h_{95} \to b\overline{b})}{{\rm BR^{SM}} (H \to b\overline{b})\vert_{ {m_{H} = m_{h_{95}}}}}
\, .
\ee
The LO cross section for the Higgs-strahlung process can be given as \cite{Kilian:1995tr}
\begin{equation}
\sigma\left(e^{+} e^{-} \rightarrow Z h_{95}\right)= \cos^2 \theta_1 \frac{G_F^2 m_Z^4}{96 \pi s}\left( v_e^2 + a_e^2\right)  \frac{\left(\lambda+12 m_Z^2 / s \right) }{\left(1-m_Z^2 / s\right)^2} 
\sqrt\lambda
\, , 
\end{equation}
where $\sqrt{s}$ is the center-of-mass energy, $\lambda = (1- (m_{h_{95}} + m_Z)^2/s) (1- (m_{h_{95}} - m_Z)^2/s)$ is the two-particle phase space function, and $v_e$ ($a_e$) is the vector (axial-vector) coupling of $Z e^+ e^-$ vertex.    
Ignoring the mixing effects between the neutral gauge bosons given in (\ref{MsqZs}), the coupling $Z e^+ e^-$ is the same as SM, {\it i.e.} $a_e = -1$ and $v_e = -1 + 4 \sin^2 \theta_W$, and hence one can obtain
\be
\mu_{bb, {\rm LO}} \simeq \cos^2\theta_1  \times \frac{{\rm BR} (h_{95} \to b\overline{b})}{{\rm BR^{SM}} (H \to b\overline{b})\vert_{ {m_{H} = m_{h_{95}}}}} \, ,
\ee
where radiative corrections known up to now \cite{Choi:2021nql} for the branching ratios of $h_{95} \to b\bar{b}$ and $H \to b\bar{b}$ are included.  
We emphasize that 
one-loop electroweak radiative corrections
to $e^-e^+ \rightarrow ZH$ have been computed within the standard model~\cite{Fleischer:1982af,Denner:1992bc,Kniehl:1991hk,Sun:2016bel,Bondarenko:2018sgg} and several beyond standard models~\cite{Heng:2015fka,Abouabid:2020eik,Aiko:2021nkb}.  At LEP center-of-mass energy regions ($\sqrt{s} \sim 200$ GeV), full one-loop electroweak corrections are about $\sim -20\%$ contributions. However, the corrections are mainly from initial-state radiative (ISR) corrections (about $\sim -18\%$ contributions as indicated in~\cite{Greco:2017fkb,Bondarenko:2018sgg}). We know that the ISR corrections are universal for many processes. Therefore, the corrections  are cancelled out in the signal strength given in (\ref{signalh1bb}).  For the reasons explained above,  it is enough to take tree-level cross sections for the processes $e^-e^+ \rightarrow ZH, Z h_{95}$ in our analysis.

\section{Numerical results}
\label{sec:Numbers}

\begin{figure}[htbp!]
        \centering
	\includegraphics[width=0.45\textwidth]{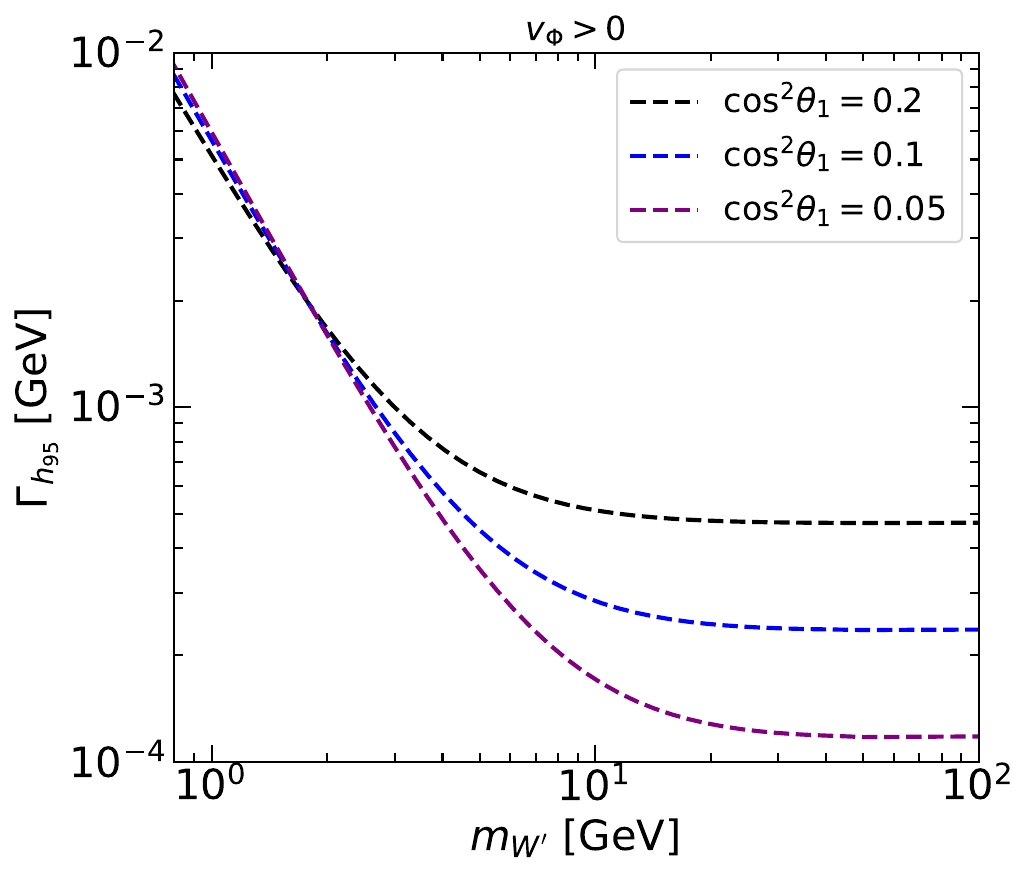}
	\includegraphics[width=0.45\textwidth]{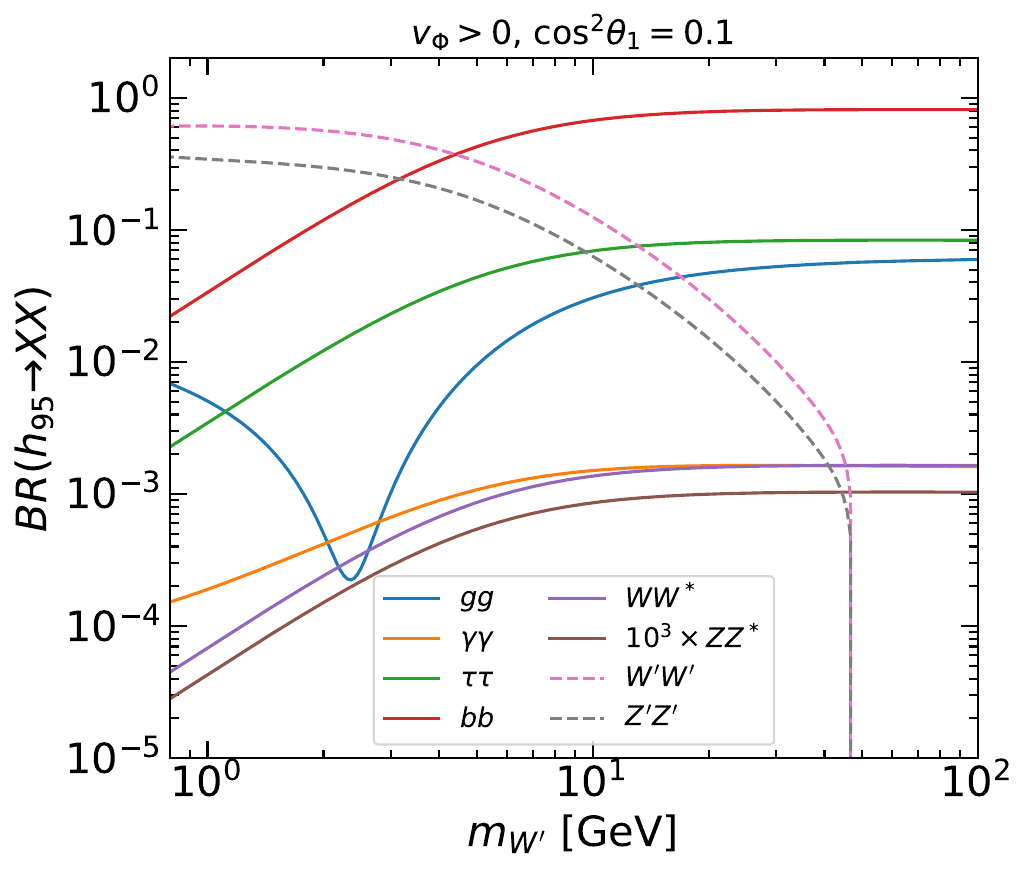}
        \includegraphics[width=0.45\textwidth]{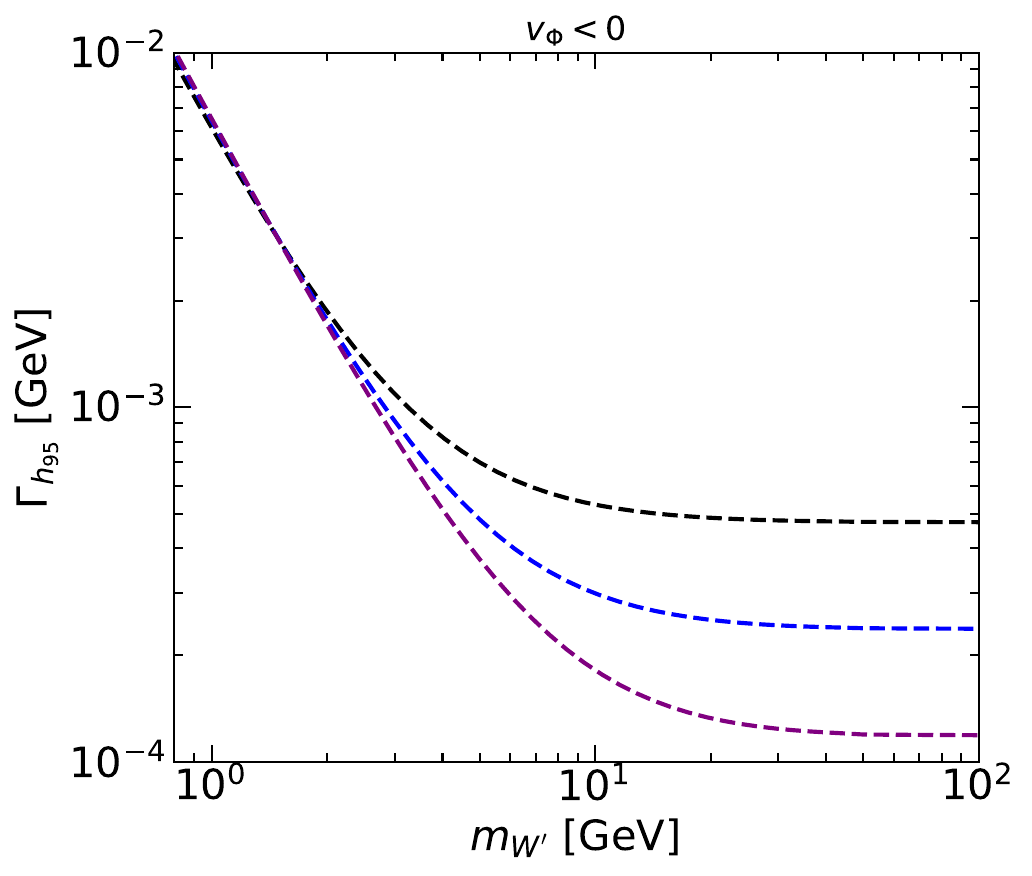}
	\includegraphics[width=0.45\textwidth]{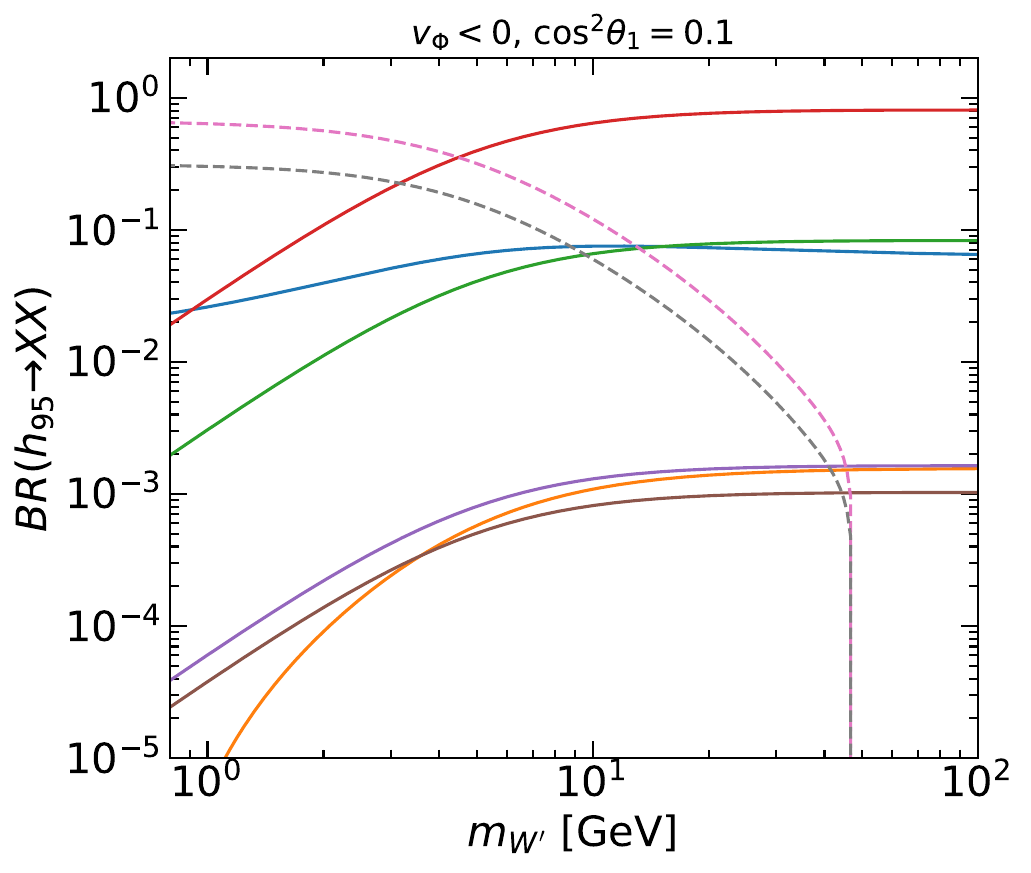}
	\caption{ \label{fig:h95_decayrate_BRs} The decay rate and branching ratio of $h_{95}$ boson as a function of DM mass with $v_\Phi > 0$ (top panels) and $v_\Phi < 0$ (bottom panels). Here we fixed $m_{h_{95}} = 95$ GeV, $m_{H^\pm} = 600$ GeV, $m_D = 550$ GeV, $g_H = 10^{-3}$, $g_X = 10^{-6}$, $m_X = 1$ TeV and $y_{f^H} = 1$. Left panels: dashed lines in black, blue, and purple denote decay rates corresponding to $\cos^2\theta_1$ values of 0.2, 0.1, and 0.05, respectively. Right panels: We fixed $\cos^2\theta_1 = 0.1$. Solid lines in blue, orange, green, red, brown and purple indicate the branching ratios of the $h_{95}$ boson decaying into pairs of SM particles, including gluons, photons, tau leptons, bottom quarks, $Z$ and $W$ bosons, respectively. Here, the branching ratio to a pair of $ZZ^*$ bosons (solid brown line) is amplified by a factor of $10^{3}$. Meanwhile, dashed lines in brown and pink represent pairs of BSM particles, $W'$ and $Z'$ respectively. }
\end{figure}

\subsection{Decay Rates and Branching Ratios of the Lighter Scalar $h_{95}$}
If kinematically allowed, the lighter scalar $h_{95}$ may decay into a pair of SM particles, including fermions and gauge bosons, and a pair of dark sector particles such as DMs, dark photons, and dark Z bosons.

Fig.~\ref{fig:h95_decayrate_BRs} illustrates decay rate and branching ratio of $h_{95}$ boson as a function of DM mass.
Here we fixed $m_{h_{95}} = 95$ GeV, $m_{H^\pm} = 600$ GeV, $m_D = 550$ GeV, $g_H = 10^{-3}$, $g_X = 10^{-6}$, $m_X = 1$ TeV and $y_{f^H} = 1$.
In the top panels, we fix the VEV $v_\Phi$ to be positive, whereas in the bottom panels, it is set to be negative. 
From the left panels of Fig.~\ref{fig:h95_decayrate_BRs}, we observe that the decay rate of the $h_{95}$ boson is notably depended on the DM mass, particularly in the low mass region ($m_{W'} < m_{h_{95}}/2$), where decaying into pairs of DMs and $Z'$ bosons becomes kinematically feasible. Furthermore, the dependence of the decay rate on the mixing angle $\theta_1$ is significant in the high DM mass region but less pronounced in the low DM mass region. 

From the right panels of Fig.~\ref{fig:h95_decayrate_BRs}, one sees that the branching ratios of $h_{95}$ to pairs of DMs and to $Z'$ bosons are dominant in the low DM mass region while higher DM mass region the branching ratios to SM particles becomes predominant. This implies that in a heavy DM mass region, there is a relatively large signal strength for the decay of $h_{95}$ into SM particles, such as di-photon and di-tau, as observed at the LHC.
Notably, the branching ratios to pairs of gluons and photons undergo significant alterations in the low DM mass region, depending on the sign of $v_\Phi$. This effect arises from the interference between contributions from charged Higgs, SM quarks and hidden quarks for the process of $h_{95}$ decaying into di-photon. Whereas for the process of $h_{95}$ decaying into gluons, it is due to the interference between contributions from SM quarks and hidden quarks. 
Note that these alterations in the decays into pairs of gluons and photons result in slight changes in the total decay width of $h_{95}$ in the low mass region for different signs of $v_\Phi$ as shown in the left panels of Fig.~\ref{fig:h95_decayrate_BRs}. 
For the decay of $h_{95}$ to a pair of the SM vector bosons  $VV^*$ (where  $V \equiv W, Z$), one of the vector bosons is necessitated to go off-shell.

\subsection{Correlations between $h_{95, 125} \to \gamma \gamma$ and $h_{95, 125} \to Z \gamma$}

In this section, we present an analysis of the correlations between the modes $h_{95, 125} \rightarrow \gamma\gamma$ and $h_{95, 125} \rightarrow Z \gamma$. These processes occur via one-loop induced with the $W^\pm$ boson, SM fermions, hidden fermions, and charged Higgs running in the loop. We find that the contribution from the hidden fermion loop is negligible for our parameters of interest.

\begin{figure}[t!]
        \centering
	\includegraphics[width=0.45\textwidth]{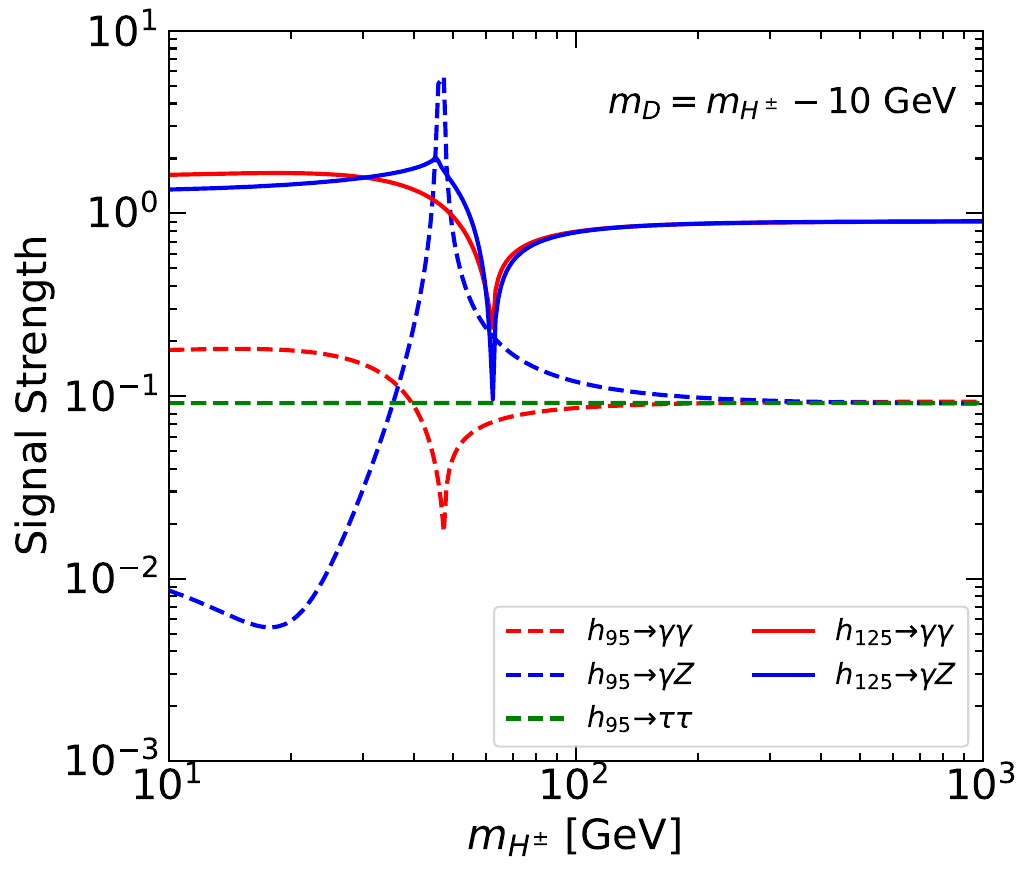}
	\includegraphics[width=0.45\textwidth]{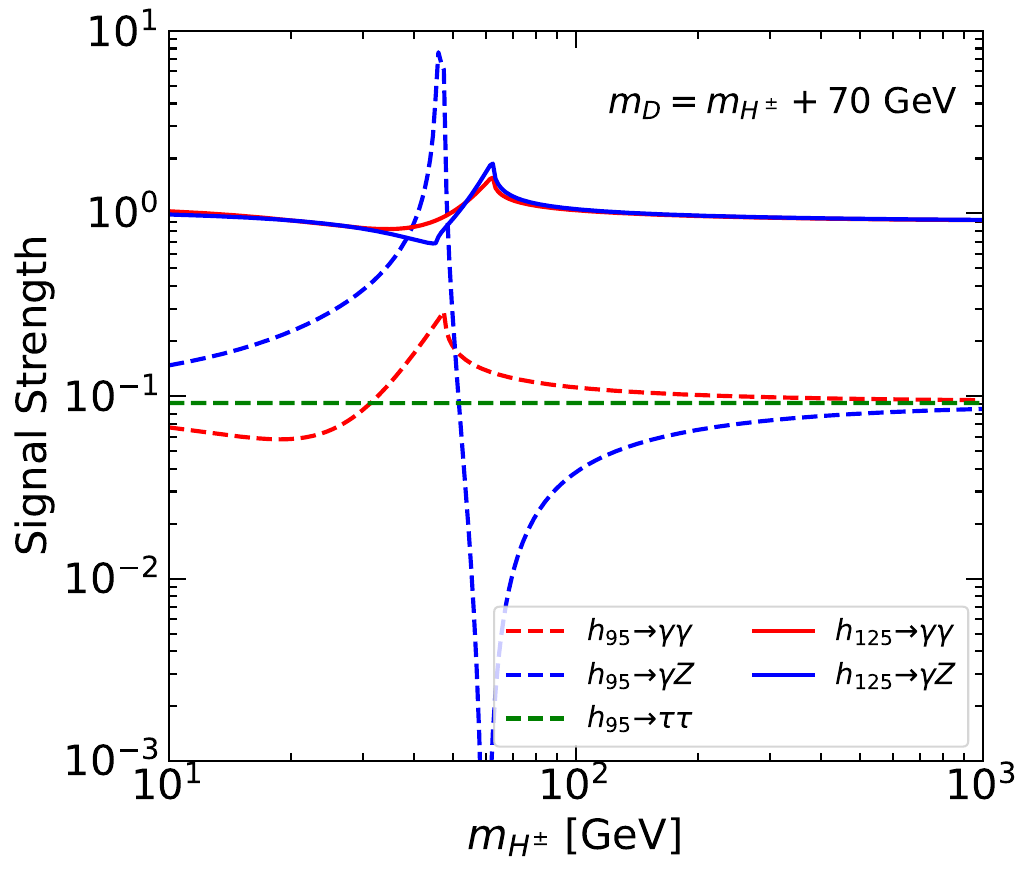}
	\caption{\label{fig:ss_mHpm} The signal strengths of lighter scalar and SM 125 GeV Higgs bosons as a function of charged Higgs mass. Here we set $m_D = m_{H^\pm} - 10$ GeV in the left panel and $m_D = m_{H^\pm} + 70$ GeV in the right panel. The remain parameters are set to be $m_{h_1} = 95$ GeV, $\cos^2\theta_1 = 0.1$, $m_{W'} = 50$ GeV, $g_H = 10^{-3}$, $g_X = 10^{-6}$, $m_X = 1$ TeV and $y_{f^H} = 1$ in both panels. The dashed red, blue and green lines represent the signal strength of lighter scalar boson decays into di-photon, $Z \gamma$ and di-tau, respectively. Whereas the solid red and blue lines indicate the signal strength of SM 125 GeV Higgs boson decays into di-photon and $Z \gamma$ , respectively. }
\end{figure}

In Fig.~\ref{fig:ss_mHpm}, we illustrate the signal strength of both lighter scalar $h_{95}$ and Higgs bosons $h_{125}$ decaying into di-photon, $Z\gamma$, and di-tau as a function of the charged Higgs mass. Here, for both scalar bosons, the production mode considered is gluon-gluon fusion.
With parameters set as follows: $m_{h_1} = 95$ GeV, $\cos^2\theta_1 = 0.1$, $m_{W'} = 50$ GeV, $g_H = 10^{-3}$, $g_X = 10^{-6}$, $m_X = 1$ TeV, and $y_{f^H} = 1$, the signal strength of $h_{95} \to \tau^+ \tau^-$ is approximately 0.1.
The signal strengths of the di-photon and $Z\gamma$ final states undergo significant alterations in the low mass range of the charged Higgs, where the contribution from charged Higgs loop becomes important.
Particularly noteworthy is when the charged Higgs becomes on-shell $(m_{H^\pm} < m_{h_{95, 125}}/2)$, the amplitude from the charged Higgs loop process  acquires an imaginary part, while its real part peaks at the mass threshold $(m_{H^\pm} = m_{h_{95, 125}}/2)$. 

On the other hand, while the sum of loop form factors from $W^\pm$ boson and SM fermions diagrams yields a negative value in the decay channels $h_{125} \to \gamma \gamma$ and $ h_{125} \to Z \gamma$, as well as $h_{95} \to \gamma \gamma$ processes, it exhibits a positive value in the $h_{95} \to Z \gamma$ process. 
This is mainly due to the sign changing of the $W^\pm$ boson loop form factor in the $ h \to Z \gamma$ process (shown in~(\ref{FWloopZg})) at scalar mass $m_h \sim 100$ GeV.
Depending on the relative sign between the loop form factors of the charged Higgs and the total $W^\pm$ and SM fermions contributions, the decay rate will either be enhanced or suppressed.

Here, we note that the bound on the charged Higgs mass from LEP \cite{L3:2003jyb} ($m_{H^{\pm}} > 80$ GeV) within the framework of the well-known two-Higgs-doublet models may not be directly applicable to the charged Higgs in this model. This is because of the different production and decay modes of the charged Higgs boson in this model compared to the conventional models.
Specifically, the charged Higgs boson in this model is odd under $h$-parity, necessitating its decay into a $h$-parity odd particle and a $h$-parity even particle. 
For example, charged Higgs can decay into $W^{\pm} W'$ and/or $W^{\pm}D$ followed by $D \to W'h_i$ and/or $D \to W'Z_i$. 
Nevertheless, searches for multilepton or multijet plus missing energy events at the LHC can establish constraints on the charged Higgs mass, resembling signatures similar to searches for charginos and neutralinos in supersymmetry \cite{ATLAS:2021moa, CMS:2021cox}.  
Moreover, one can put a lower bound on the charged Higgs mass in minimal G2HDM depending upon the hidden up-type quark mass using the data from rare $B$ meson decays and oblique parameters \cite{Liu:2024nkl}. 
Taking into account these considerations, in scanning results presented in following subsections, we set $m_{H^{\pm}} > 100$ GeV.

In the left panel of Fig.~\ref{fig:ss_mHpm}, we set $m_D = m_{H^\pm} - 10$ GeV, ensuring that the coupling between $h_{95, 125}$ and charged Higgs (as shown in \ref{ghiCHCH}) is positive. Consequently, a cancellation between the loop contributions causes dips at the mass thresholds for the signal strength of $h_{125} \to \gamma \gamma$ and $Z \gamma$, as well as $h_{95} \to \gamma \gamma$, while an enhancement between the loop contributions results in a peak shape for the signal strength of $h_{95} \to Z \gamma$.

On the other hand, when $m_D = m_{H^\pm} + 70$ GeV is assumed, the coupling between $h_{95, 125}$ and charged Higgs becomes negative, resulting in peaks at the mass thresholds for the signal strength of $h_{125} \to \gamma \gamma$, $h_{125} \to Z \gamma$ and $h_{95} \to \gamma \gamma$ as shown in the right panel of Fig.~\ref{fig:ss_mHpm}. In the same panel, for the $h_{95} \to Z \gamma$ signal strength, there is enhancement at charged Higgs masses below the mass threshold due to the dominance of the imaginary part from the charged Higgs loop contribution, while there is suppression above the mass threshold due to the cancellation between real part of the charged Higgs, $W^\pm$, and SM fermion loop contributions.

From Fig.~\ref{fig:ss_mHpm}, we find a strong correlation between the signal strengths of $h_{125} \to \gamma \gamma$ and $h_{125} \to Z \gamma$, whereas this correlation does not extend to the lighter scalar boson $h_{95}$. This feature persists even after conducting a comprehensive parameter scan, as presented in the following subsection.
Moreover, with $\cos^2 \theta_1 = 0.1$, the signal strength of $h_{95} \to \gamma \gamma$ is approximately one order of magnitude smaller than the same channel for $h_{125}$. A larger value of $\cos^2 \theta_1$ results in an increase in signal strengths of $h_{95}$ and a decrease in $h_{125}$.

We note that we are focusing on $h_{95,125}\to \gamma\gamma, Z\gamma$ in this work. The other one-loop induced processes $h_{95, 125}\to \gamma^\prime \gamma, Z^\prime \gamma$, {\it i.e.} $h_i \to Z_{2,3} \gamma$, if kinematically allowed~\footnote{In the event that $m_{h_i} < m_{Z^\prime}$, we can consider the reverse process $Z^\prime \to h_i \gamma$.}, are also of potential interest. However their rates in minimal G2HDM are roughly suppressed by $(g_X/g)^2$ 
and $(g_H/g)^2$ as compared to the case of $h_i \to Z \gamma$. We will leave them for future pursuits.

\subsection{Scanning Results}

We scan over the parameter space in the model through all
constraints discussed in Sec.~\ref{sec:constraints}. 
We note that the detailed discussion on the constraints on the model have been studied in previous works~\cite{Ramos:2021txu, Ramos:2021omo}.
 
We employ the \textit{Profile Likelihood} method~\cite{Rolke:2004mj} to eliminate nuisance parameters and project the results on two dimensional plane. 
For theoretical bounds and dark photon constraints, we apply a hard cut, meaning any points outside the bounds are simply discarded. 
To incorporate constraints from the Higgs signal strength at CMS, electroweak precision measurements at the $Z$ pole, the $W$ boson mass, and the dark matter relic density, we compute the chi-square function of each constraint as follows:
\be
    \chi^2 = \left( \frac{O_{\rm th} - O_{\rm exp}}{\sigma} \right)^2 ~ \text{with} ~
    \sigma = \sqrt{\sigma^2_\mathrm{th} + \sigma^2_\mathrm{exp}} \; ,
\ee
where $O_{\rm th}$ represents the theoretical prediction, $O_{\rm exp}$ denotes the experimental central value, and $\sigma_\mathrm{th}$ ($\sigma_\mathrm{exp}$) represents the theoretical (experimental) uncertainty.
For the DM direct detection, Higgs invisible decays and the $Z'$ high-resonance limits, reported at $95\%$ C.L. under the assumption of a null signal, 
we use the following chi-square function for each of these constraints: 
\be
\chi^2 = 2.71 \times \left( \frac{O_{\rm th}}{O_{\rm exp}^{\rm limit}}\right)^2, 
\ee
where $O^{\mathrm{limit}}_\mathrm{exp}$ is the $95\%$ C.L. experimental limit.
The total chi-square, $\chi^2_{\rm total}$, is obtained by summing the individual chi-square value from each constraint.

To sample the parameter space in the model, we employ MCMC scans using {\tt emcee}~\cite{ForemanMackey:2012ig} with a log-likelihood function, $\log{\cal L} = -0.5 \times \chi^2_{\rm total}$. The scan range is given as, 
\bea
m_{h_{95}}/ {\rm GeV} &\in& (94,\, 97) \,, \\
m_{H^{\pm}} /{\rm GeV}  &\in& (100,\, 1000) \,, \\
m_{D} /{\rm GeV}  &\in& (10,\, 1000) \,, \\
m_{W'}/{\rm GeV}&\in& (0.1,\, 1000) \, , \\
\theta_{1}/{\rm rad}  &\in& \left(-\frac{\pi}{2},\, \frac{\pi}{2} \right) \,, \\
v_{\Phi}/{\rm GeV}&\in& (10^2,\, 10^{5}) \, , \\
g_X  &\in& (10^{-6},\, 10^{-2}) \,, \\
m_{X}/{\rm GeV} &\in& (10^{-2},\, 10^{3})  \,,
\eea
and we fix the Yukawa couplings of the hidden fermions to be $y_{f^H} = 1$.
The parameters $m_{W'}$, $v_{\Phi}$, $g_X$ and $m_{X}$ are sampling in log prior while the remain ones are in linear. 
We obtain approximately $10^5$ data points. The confidence intervals are derived from the tabulated values of $\Delta\chi^2 \equiv -2\log(\mathcal{L/L}_{\rm max})$, where $\mathcal{L}_{\rm max}$ represents the likelihood at the best-fit value. For a two-dimensional plot, the $95\%$ C.L. region (the $2\sigma$ allowed region) is defined by $\Delta\chi^2 \leq 5.99$, assuming an approximately Gaussian likelihood.

\begin{figure}[t!]
        \centering
	\includegraphics[width=0.45\textwidth]{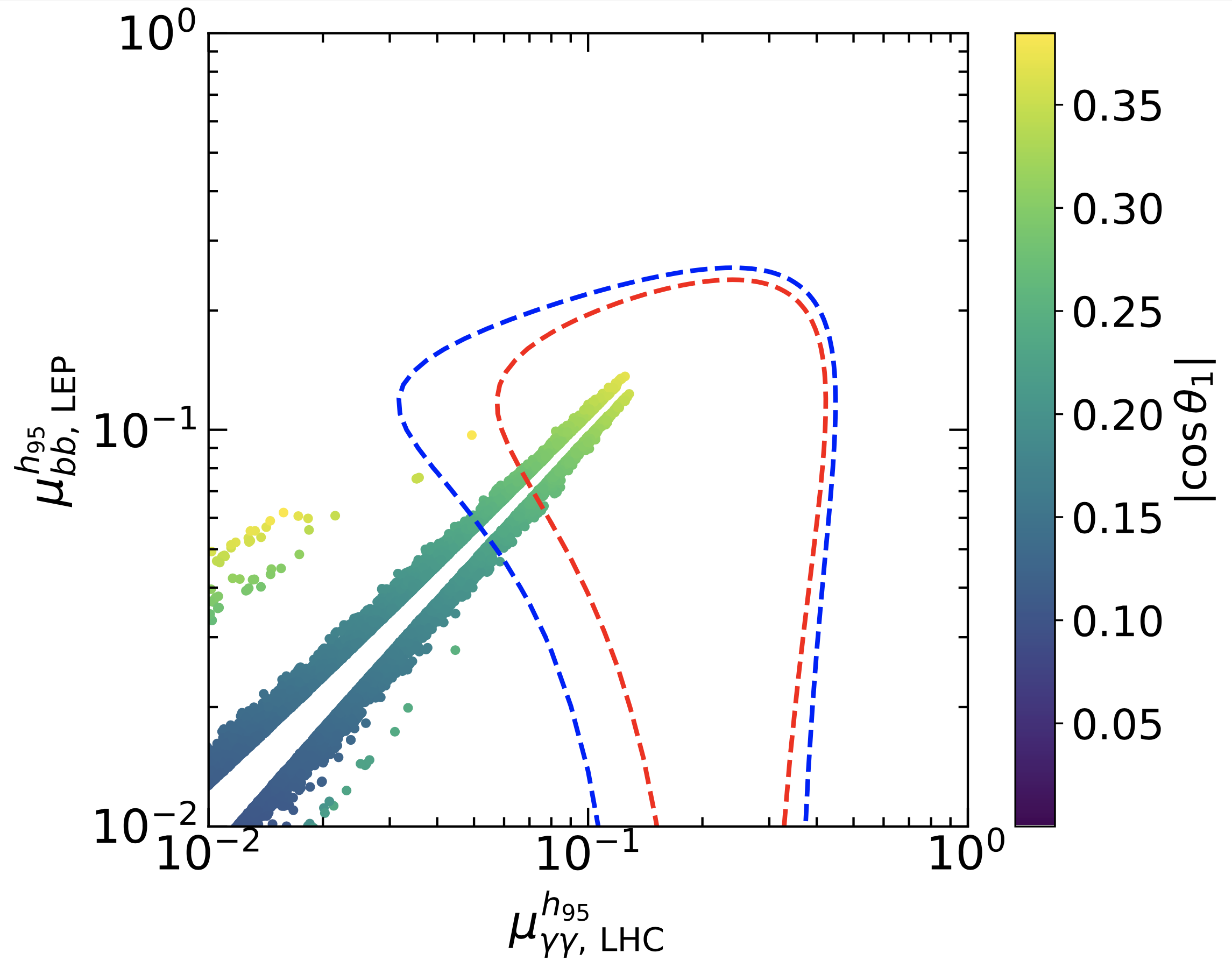}
	\includegraphics[width=0.45\textwidth]{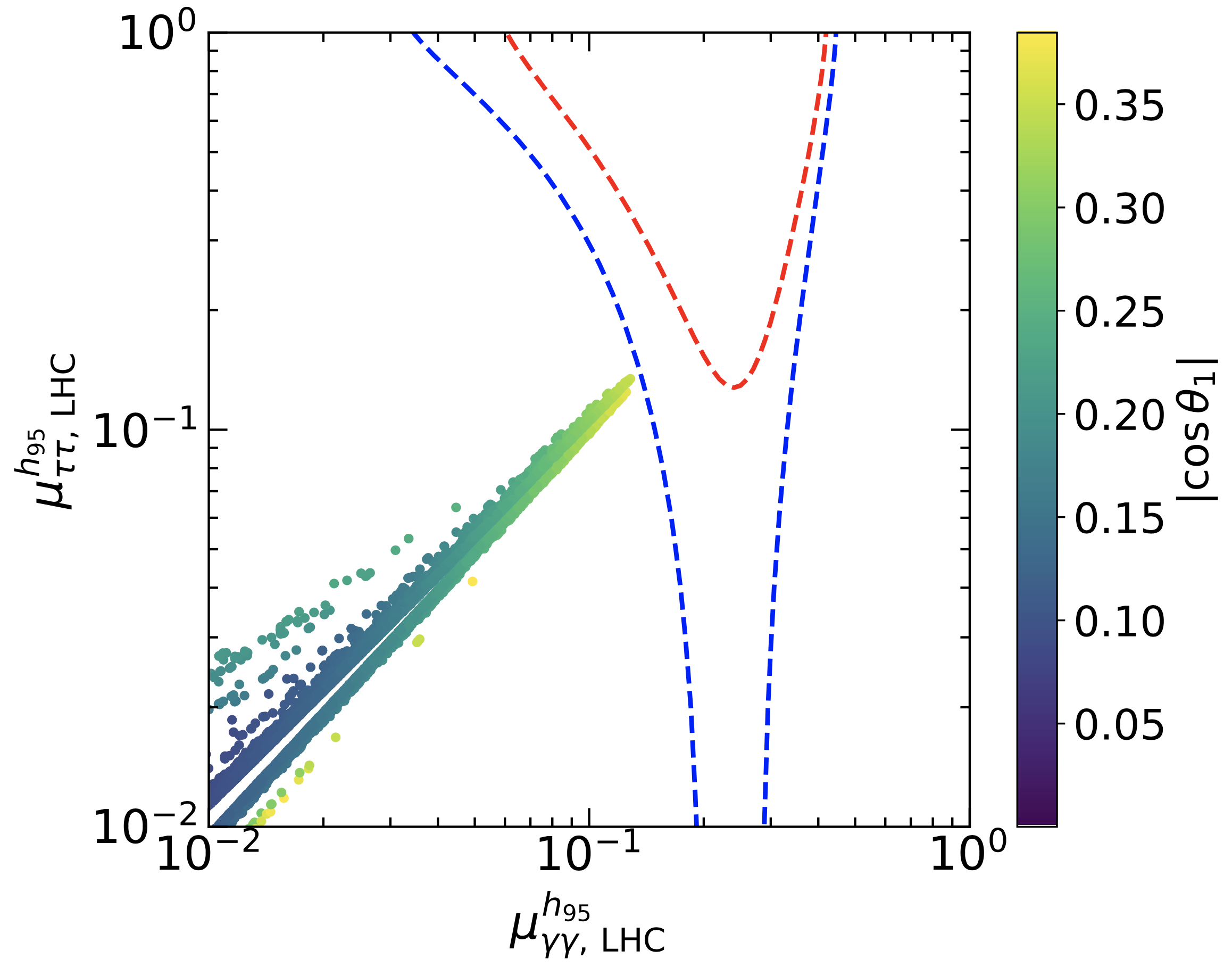}
        \includegraphics[width=0.45\textwidth]{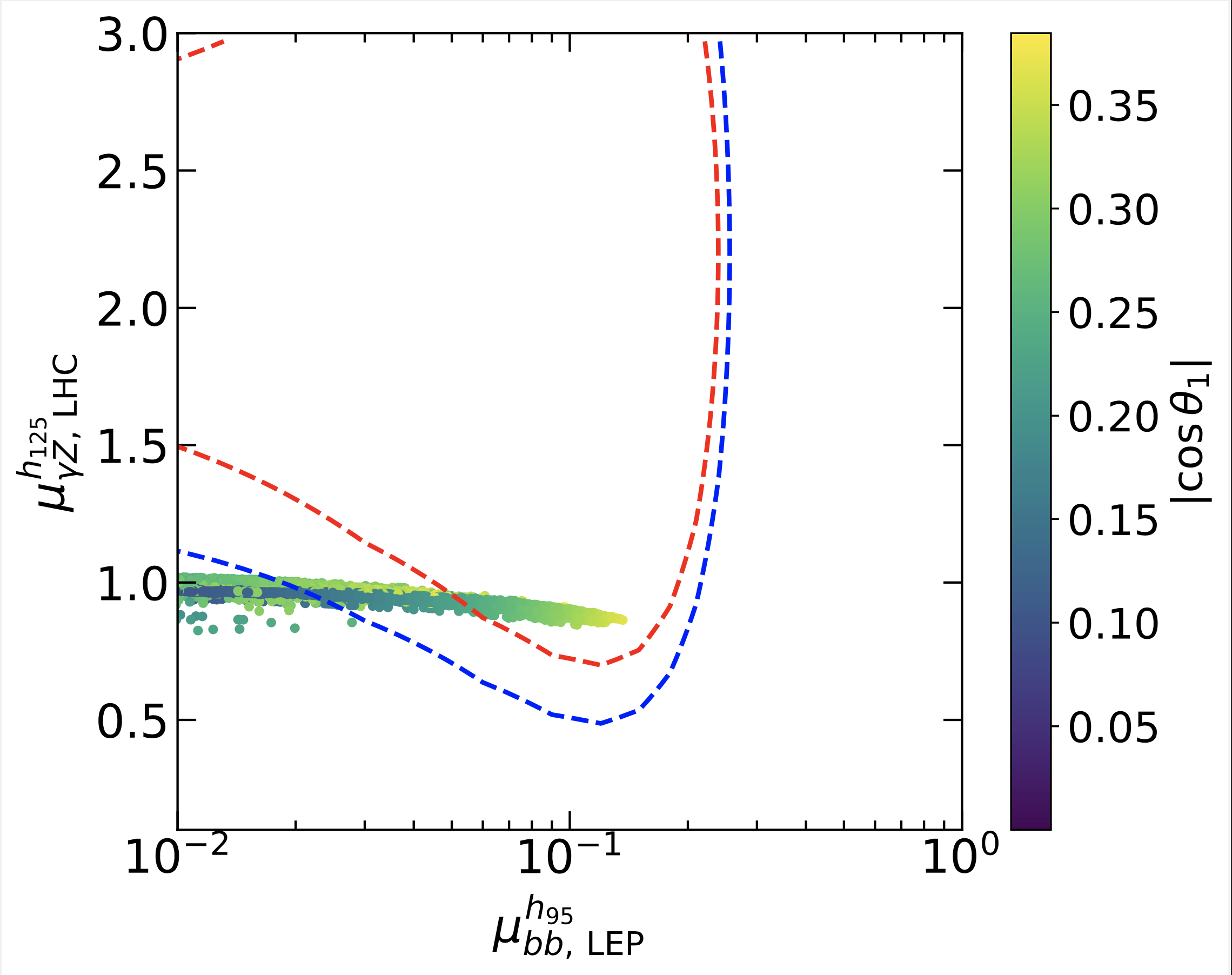}
	\includegraphics[width=0.45\textwidth]{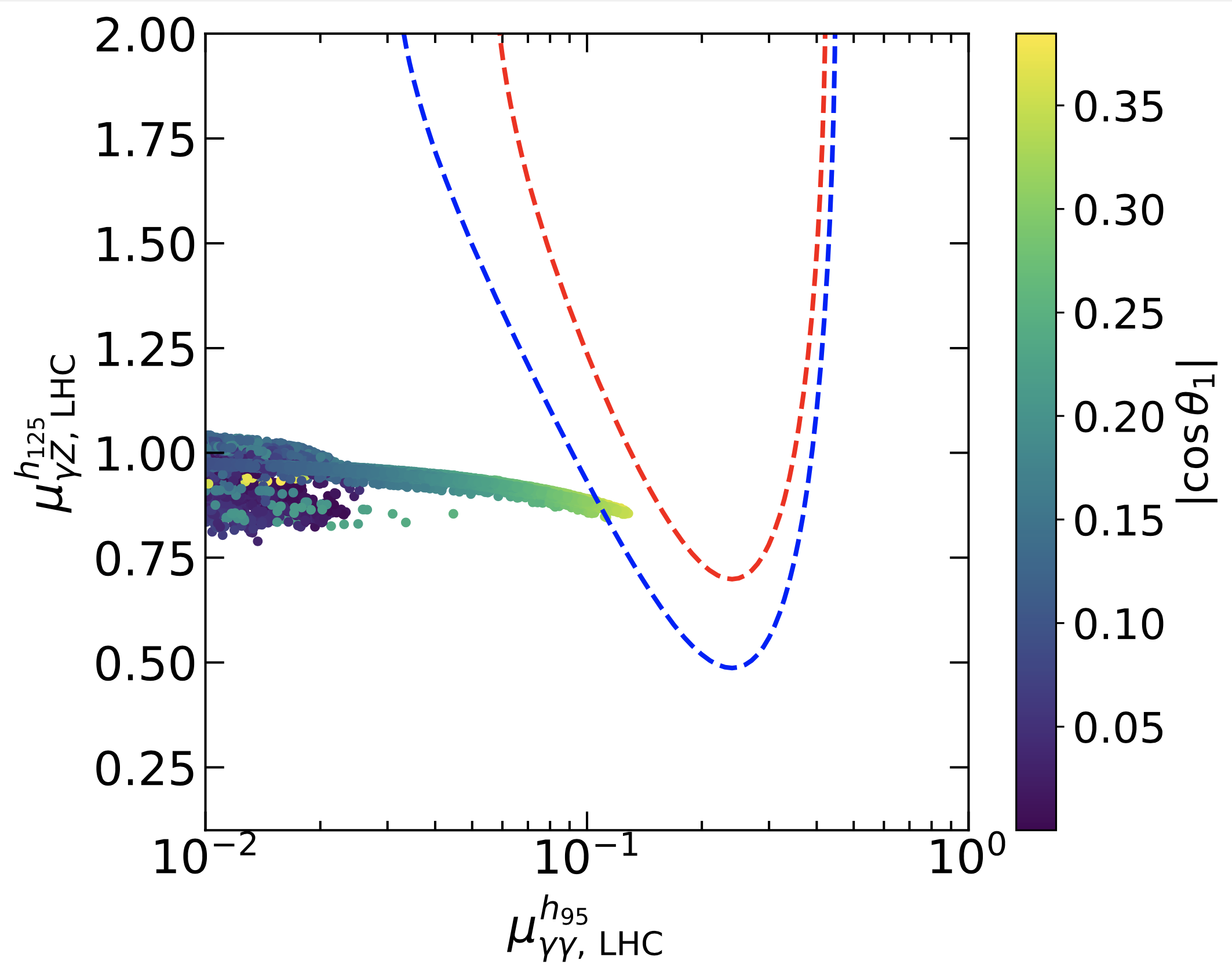}
	\caption{\label{fig:mu} The signal strength of lighter scalar boson $h_{95}$ and SM Higgs boson $h_{125}$. The red and blue lines represent $1\sigma$ and $2\sigma$ experimental contours, respectively. From left to right and top to bottom, the viable parameter space projected on ($\mu_{\gamma \gamma}^{h_{95}}$, $\mu_{bb}^{h_{95}}$), ($\mu_{\gamma \gamma}^{h_{95}}$, $\mu_{\tau\tau}^{h_{95}}$), ($\mu_{bb}^{h_{95}}$, $\mu_{\gamma Z}^{h_{125}}$), ($\mu_{\gamma \gamma}^{h_{95}}$, $\mu_{\gamma Z}^{h_{125}}$) planes, respectively. The color indicates the values of $|\cos(\theta_1)|$.}
\end{figure}

Fig.~\ref{fig:mu} illustrates the viable parameter space within the model which has passed all
constraints mentioned in Sec.~\ref{sec:constraints}, projected onto the signal strength of the 95 GeV scalar boson and the SM-like scalar boson. We find that a portion of the viable data points can explain the signal strength of the 95 GeV scalar boson decaying into di-photon events, as measured at the LHC, and into $b\bar{b}$ pairs, as measured at LEP, as depicted in the top-left panel of Fig.~\ref{fig:mu}. However, the signal strength of the 95 GeV scalar boson decaying into di-tau pairs is found to be too small, thus conflicting with current measurements at CMS, as shown in the top-right panel of Fig.~\ref{fig:mu}. Notably, a larger value of $|\cos(\theta_1)|$ can result in a higher value of $\mu_{\tau\tau}^{h_{95}}$, however the current constraints from Higgs data measurements at the CMS \cite{CMS-PAS-HIG-19-005} require  $|\cos(\theta_1)| \lesssim 0.36$.

The bottom panels in Fig.~\ref{fig:mu} present the viable data points projected on planes of the signal strengths of $h_{125} \to \gamma Z$ and $h_{95} \to b\bar{b}$ (left panel), and $h_{95} \to \gamma \gamma$ (right panel). The results indicate that, within $2\sigma$ region, the model can simultaneously accommodate the experimental data for 95 GeV scalar boson decaying into $b\bar{b}$ from LEP and its decaying to di-photon from the LHC as well as the SM-like Higgs boson decaying into $Z \gamma$ channel from recent measurements at the LHC.

\begin{figure}[htbp!]
        \centering
	\includegraphics[width=0.45\textwidth]{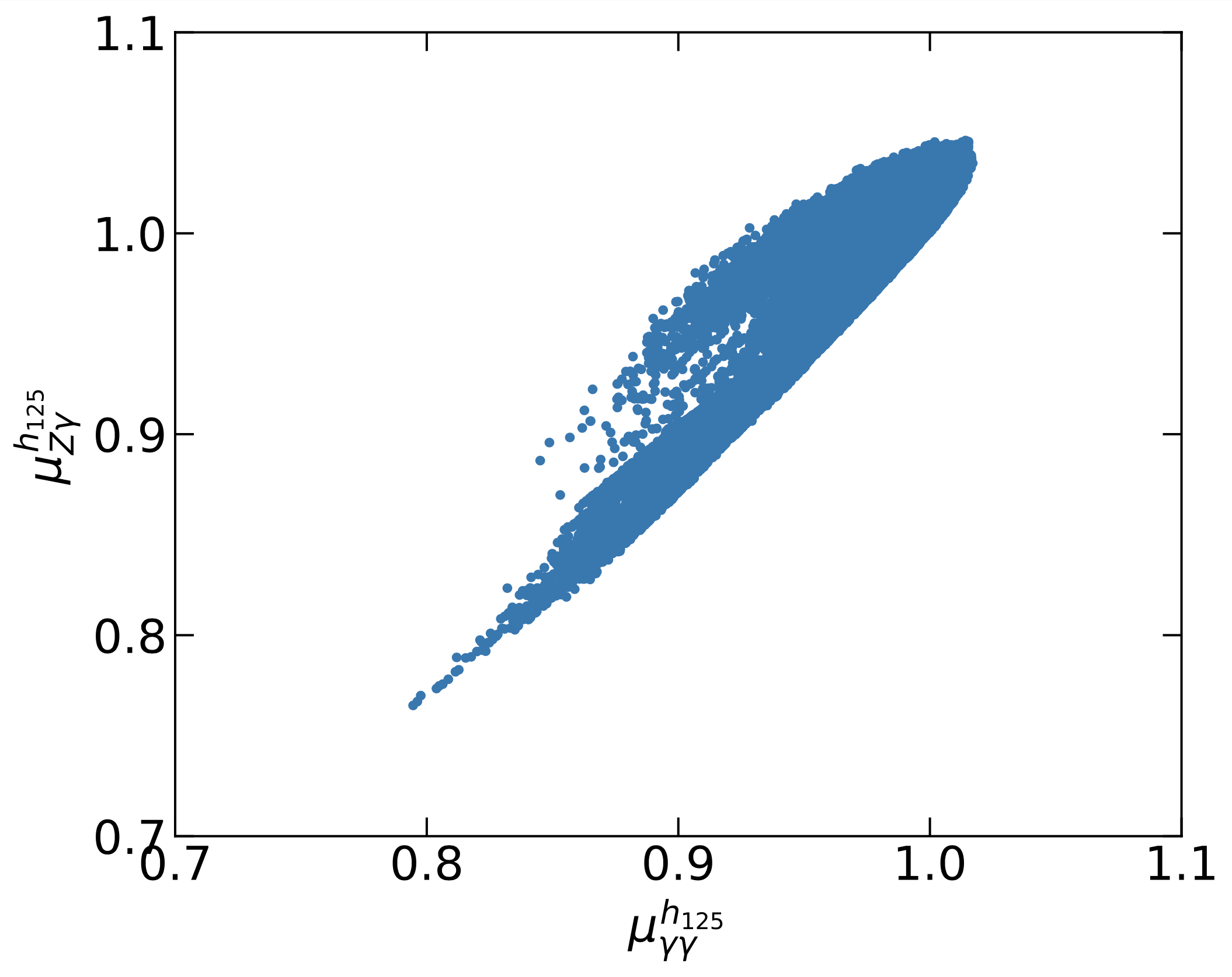}
	\includegraphics[width=0.45\textwidth]{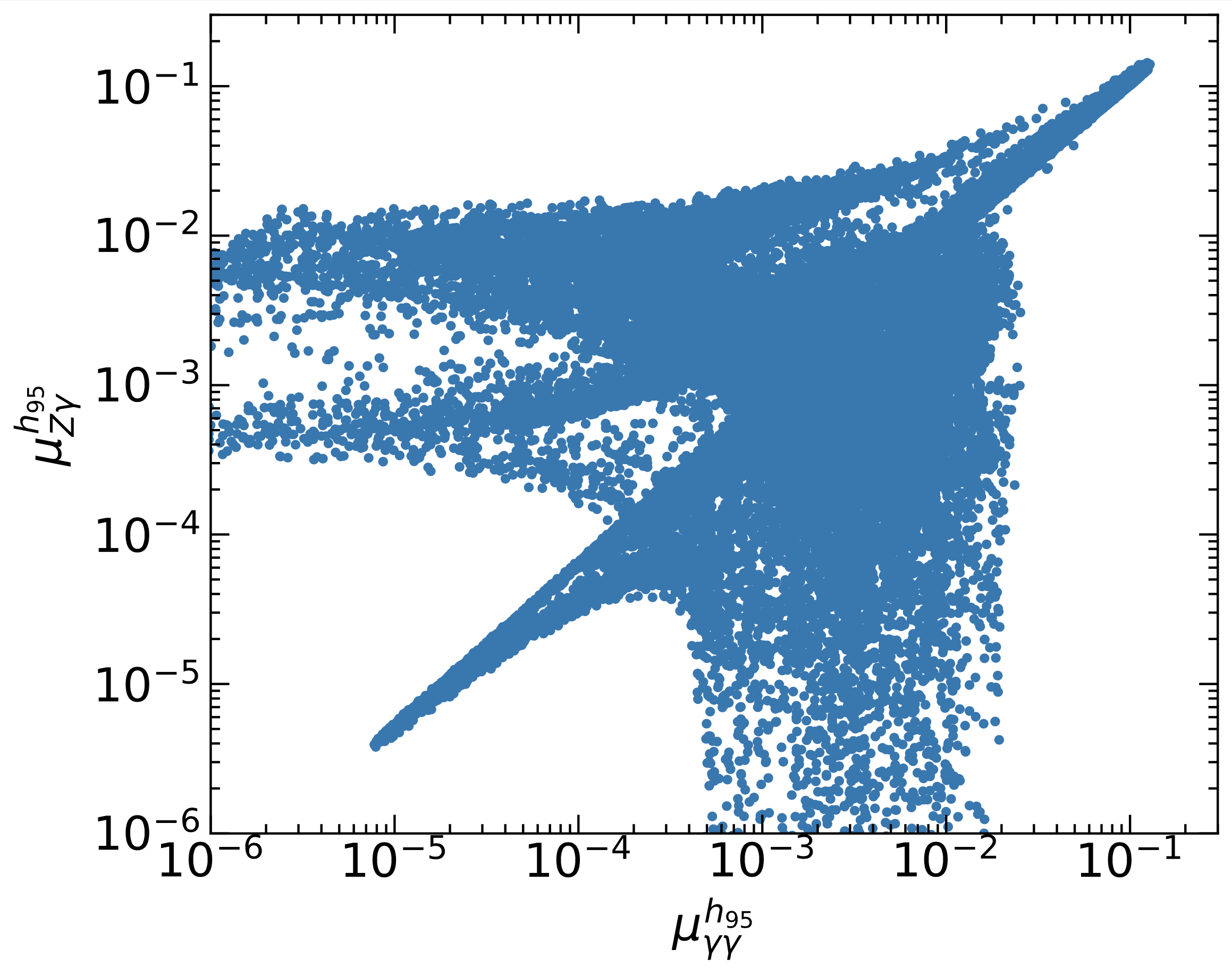}
	\caption{\label{fig:gaga_gaZ_correlation}Viable parameter space projected on the signal strengths of $h_{125} \to \gamma \gamma$ and $h_{125} \to Z \gamma$ plane (left), and $h_{95} \to \gamma \gamma$ and $h_{95} \to Z \gamma$ plane (right).}
\end{figure}

In Figure~\ref{fig:gaga_gaZ_correlation}, we present the detailed correlation between signal strengths of $h_{95, 125} \to \gamma \gamma$ and $h_{95, 125} \to Z \gamma$ within the viable parameter space.
In the left panel of Fig.~\ref{fig:gaga_gaZ_correlation}, we show the viable parameter space spanned on the signal strengths of $h_{125} \to \gamma \gamma$ and $h_{125} \to Z \gamma$ plane. The signal strengths of $h_{95} \to \gamma \gamma$ and $h_{95} \to Z \gamma$ is presented in the right panel. 
Here, for both $h_{125}$ and $h_{95}$, the production mode considered is gluon-gluon fusion. 
It is evident that the signal strengths of $h_{125} \to \gamma \gamma$ and $h_{125} \to Z \gamma$ exhibit a strong correlation, whereas this correlation is not observed for the $h_{95}$ boson. 


\section{Conclusion}\label{sec:Conclusions}

Recent results from low-mass Higgs boson searches at the LHC reveal excesses around 95 GeV in the di-photon final state, with a local significance of 3.1$\sigma$ when combining CMS and ATLAS data, and in the di-tau final state, with a local significance of 2.6$\sigma$ from CMS. An excess at a similar mass, with a local significance of 2.3$\sigma$, was previously observed at the LEP experiment. Additionally, the LHC has reported the first evidence of a rare decay mode of the 125 GeV Higgs boson into $Z \gamma$.

In this study, we investigate the simultaneous interpretation of the excesses at 95 GeV arising from the production of a lighter scalar boson and the rare decay mode of the 125 GeV Higgs boson into $Z \gamma$ within the framework of the gauged two-Higgs-doublet model.

We have presented an analysis of the decay properties of the lighter scalar boson $h_{95}$. In addition to its decays into SM particles due to the mixing with SM Higgs boson, $h_{95}$ can decay into particles within the dark sector via its major hidden component, including DM, dark photons, and dark $Z$ bosons, where kinematically feasible. The decay rate of $h_{95}$ to SM particles is significantly influenced by the mixing angle $\theta_1$, with a larger value of $|\cos(\theta_1)|$ correlating to higher decay rates. Notably, the $h_{95} \rightarrow b\bar{b}$ channel emerges as the predominant decay mode among SM particle final states. On the other hand, the decay rate of $h_{95}$ into dark sector particles strongly depends on the new gauge couplings and the masses of the final state particles. For a substantial gauge coupling and a low mass region of DM, the branching ratios of $h_{95}$ decaying into dark sector particles can dominate, as illustrated in Fig.~\ref{fig:h95_decayrate_BRs}.

We focused our investigation on the di-photon and $Z\gamma$ final states from the decays of both 95 GeV and 125 GeV Higgs bosons. In addition to the anticipated contributions from $W^{\pm}$ boson and SM fermions loops, our analysis found significant effects from charged Higgs loop, especially in low charged Higgs mass region. The impact of the hidden fermions loop remains negligible within our parameter range of interest. The signal strengths of $h_{95, 125} \to \gamma \gamma $ and $Z\gamma$ can either be enhanced or suppressed depending on the constructive or destructive interference between these contributions, as depicted in Fig.~\ref{fig:ss_mHpm}.  

Employing an exhaustive parameter space scan, constrained by the theoretical conditions and experimental data, we present our main results in Fig.~\ref{fig:mu}. 
We found that the viable parameter space in the model can simultaneously address the excesses observed around 95 GeV in the $b\bar{b}$ final state channel at LEP and the di-photon final state channel at the LHC as well as the recent evidence for the 125 GeV Higgs boson decay into $Z\gamma$ at the LHC. 
On the other hand, the signal strength of $h_{95} \to \tau^+ \tau^-$ is insufficient to account for the excess reported by CMS. 
Moreover, within the viable parameter space, we found a strong correlation between the signal strengths of $h_{125} \to \gamma \gamma$ and $h_{125} \to Z \gamma$, although this correlation doesn't extend to the lighter scalar $h_{95}$, as shown in Fig.~\ref{fig:gaga_gaZ_correlation}.

Upcoming results from the ATLAS experiment for low-mass searches in the di-tau final state channel, along with forthcoming Run 3 results from both ATLAS and CMS, particularly in the di-photon final state channel, hold promise for shedding light on the potential presence of an additional scalar boson around 95 GeV.

\vskip 0.5in


\section*{Acknowledgments}
This work was supported in part by the Moroccan Ministry of Higher Education and Scientific Research MESRSFC and CNRST Project PPR/2015/6 (AB), National Natural Science Foundation of China, grant No. 19Z103010239 (VQT), and NSTC grant Nos. 111-2112-M-001-035, 113-2112-M-001-001 (TCY) and 112-2811-M-001-089 (VQT).

\appendix
\label{appendix}

\section{}
\label{appA}
The general analytical expressions for the one-loop amplitudes of the two processes 
$h_i \to \gamma \gamma$ and $h_i \to Z_j \gamma$~\footnote{See for example Refs.~\cite{Gunion:1989we,Gamberini:1987sv,Weiler:1988xn,Chen:2013vi,Hue:2017cph}
for the computation of this process in a variety of BSM.} in G2HDM were given in the Appendix in~\cite{Tran:2022yrh}. To make this paper self-contained, we briefly summarize these formulas here.
As mentioned in the text, $h_1 \equiv h_{95}$, $h_2 \equiv h_{125}$ and $Z_1$ are identified as the lighter scalar with mass $\sim 95$ GeV, the observed Higgs with mass $125.38 \pm 0.14$ GeV~\cite{CMS:2020xrn} and $Z$ boson with mass $91.1876 \pm 0.0021$  GeV~\cite{Zyla:2020zbs}, respectively. 

Define the following two well-known loop functions $I_1(\tau,\lambda)$ and $I_2(\tau,\lambda)$~\cite{Gunion:1989we}
\bea
 \label{I1}
I_1(\tau,\lambda) & = & \frac{\tau \lambda}{2 ( \tau - \lambda)} + \frac{\tau^2 \lambda^2}{2 ( \tau - \lambda )^2} \left[ f(\tau) - f(\lambda) \right] \nonumber \\ 
 &+& \frac{\tau^2 \lambda}{2 ( \tau - \lambda )^2} \left[ g(\tau) - g(\lambda) \right] \; , \\
I_2(\tau,\lambda) & = & - \frac{\tau \lambda}{2 ( \tau - \lambda )} \left[ f(\tau) - f(\lambda) \right] \; ,
\label{I2}
\eea 
with 
\bea
\label{fx}
f( x ) & = & \left\{ 
\begin{array}{cr}
 \left[ {\rm arcsin} (1 / \sqrt{x} ) \right]^2  \, , &   (x \geq 1) \, , \\
- \frac{1}{4} \left[ \ln \left( \eta_+ / \eta_- \right) - i \pi \right]^2 \, ,  & (x < 1) \, ; \end{array} \right. \\
g(x) & = & \left\{ 
\begin{array}{cr}
\sqrt{x - 1} \, {\rm arcsin} (1 / \sqrt{x} ) \, , &   (x \geq 1) \, , \\
\frac{1}{2} \sqrt{1 - x} \left[ \ln \left( \eta_+ / \eta_- \right) - i \pi \right] \, ,  & (x < 1) \, ; \end{array} \right.
\label{gx}
\eea
where 
\be
\eta_\pm \equiv 1 \pm \sqrt{1 - x} \; .
\ee

\begin{figure}[htbp]
        \centering
	\includegraphics[width=0.45\textwidth]{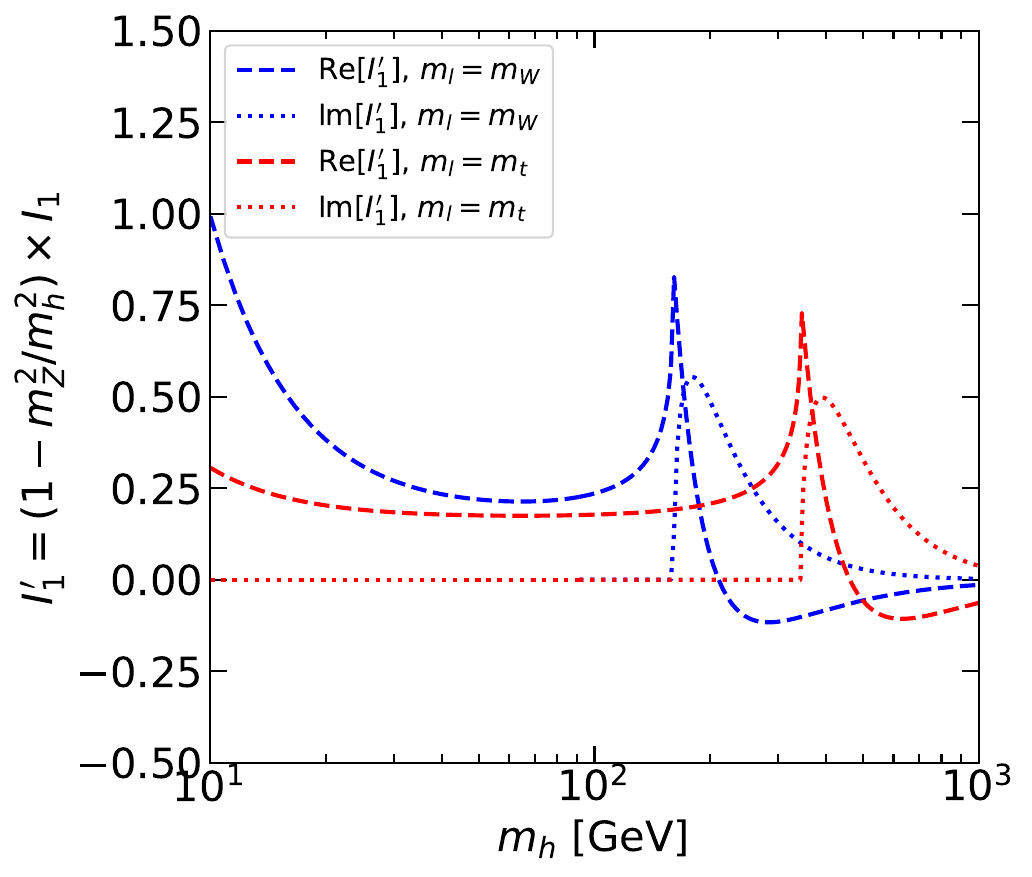}
	\includegraphics[width=0.45\textwidth]{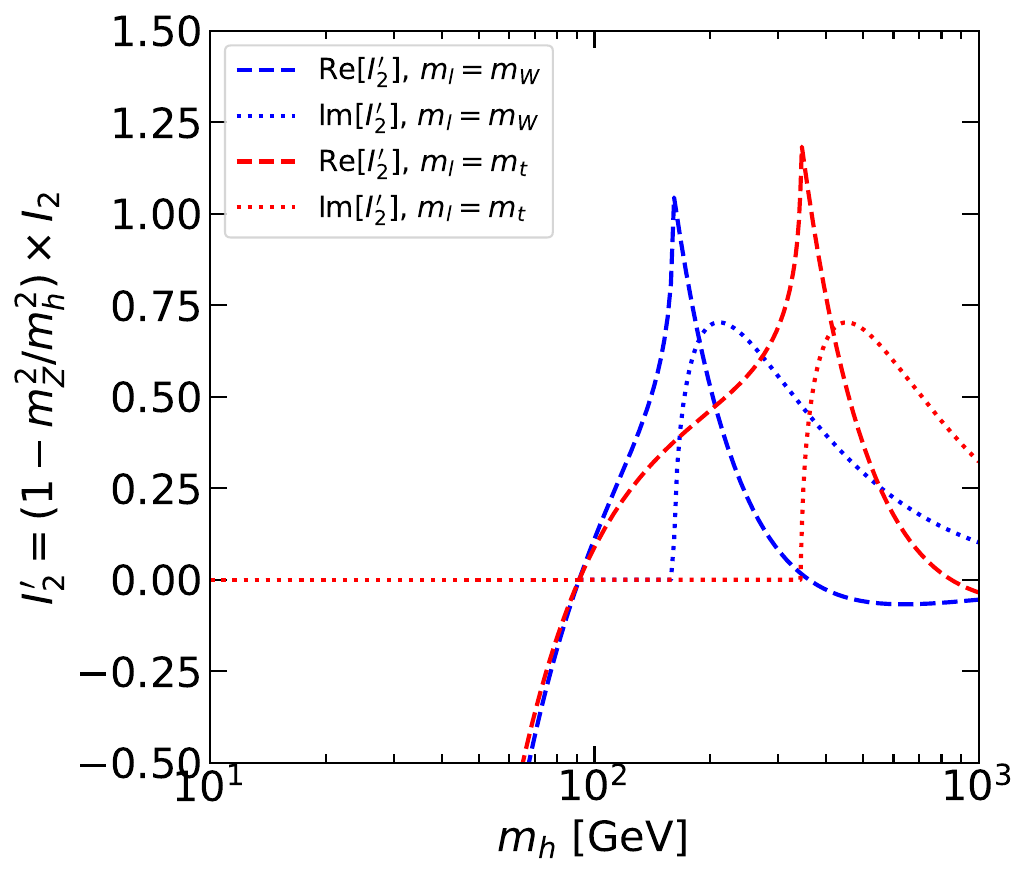}
	\caption{\label{fig:loopfunc}Loop functions $I_1^{\prime} \equiv (1 - m_Z^2/m_h^2) I_1$ (left panel) and $I_2^{\prime} \equiv (1 - m_Z^2/m_h^2) I_2$ (right panel) for $h \to Z \gamma$ process as a function of $m_h$. The dashed and dotted blue (red) represent the real and imaginary part of the loop function with mass of the particle running inside the loop $m_l = m_W$ ($m_l = m_t$).  }
\end{figure}

Fig.~\ref{fig:loopfunc} illustrates the loop functions for the $h \to Z \gamma$ process, multiplied by the phase space factor $(1 - m_Z^2/m_h^2)$, as a function of $m_h$. In this case, $\tau = 4 m_l^2/m_{h}^2$ and $\lambda = 4 m_l^2/m_{Z}^2$, where $m_l$ represents the mass of the particle running inside the loop. We note that the loop function $I_1$ exhibits a singularity at $m_h = m_Z$; however, this will be canceled out by the phase space factor. Furthermore, when the particle running inside the loop is on-shell ($m_l < m_h/2$), the loop functions acquire imaginary parts, with the real parts peaking at the mass threshold ($m_l = m_h/2$).

\subsection{Decay Rate of $h_i \to Z_j \gamma$}
\label{RatehiZjgamma}

The partial decay rate for $h_i \to Z_j  \gamma $ 
$(i =1,2; j=1,2,3)$ is 
\be
\Gamma( h_i \to Z_j \gamma ) = \frac{1}{32 \pi} m_{h_i}^3 \left( 1 - \frac{m_{Z_j}^2}{m_{h_i}^2} \right)^3 
\biggl\vert F^1_{i j }  + F^{1/2}_{i j }  + F^0_{i j }  \biggr\vert^2 \; ,
\label{RatehiZg}
\ee
where $F^s_{ij}$ with $s=0,1/2,1$ denotes the loop form factor for charged particle with spin equals $0,1/2,1$ respectively running inside the loop.

In G2HDM, the only charged spin 1 particle is the SM $W^\pm$, thus $F^1_{i j } = F_{i j } (W^\pm)$,
\bea
F_{ij} (W^\pm) & = & \frac{1}{16 \pi^2} \cdot e \cdot g m_W \cdot g c_W \cdot \frac{1}{m_W^2} \cdot {\mathcal O}^S_{1i} {\mathcal O}^G_{1j} \nonumber \\
 & & \hspace{-1.5cm} \times
\left\{
\left[ 5 + \frac{2}{ \tau_{iW}} + \left( 1 + \frac{2}{ \tau_{iW}} \right) {\left( 1 -  \frac{4}{ \lambda_{jW}} \right) } \right] I_1 \left( \tau_{iW}, \lambda_{jW} \right) \right. \nonumber \\
&  &  \left.
- 16 {\left( 1 -  \frac{1}{ \lambda_{jW}} \right)} I_2 \left( \tau_{iW}, \lambda_{jW} \right) \right\} \; .
\label{FWloopZg}
\eea

Here and below, we will denote $\tau_{il} = 4 m^2_l / m^2_{h_i}$ and $\lambda_{jl} = 4 m^2_l / m^2_{Z_j}$.

All the charged fermions in G2HDM, including both the SM fermions $f^{\rm SM}$ and the new heavy fermions $f^{H}$ contribute to 
$F^{1/2}_{ij}$. Thus
\be
F^{1/2}_{i j } = \sum_{f^{\rm SM}} F_{ij} ( f^{\rm SM} ) + \sum_{f^H} F_{ij}  ( f^H ) \; ,
\label{Fonehalf}
\ee
where
\bea
F_{ij } ( f^{\rm SM} ) & = &  \frac{1}{16 \pi^2} \cdot N^c_{ f^{\rm SM}} \cdot e Q_{ f^{\rm SM}}  \cdot \frac{m_{ f^{\rm SM}}} {v} 
 \cdot C^{f^{\rm SM}}_{Vj} \cdot \frac{-2}{m_{f^{\rm SM}}} \cdot {\mathcal O}^S_{1i} \nonumber \\
 & \, & \times \biggl[ I_1 \left( \tau_{if^{\rm SM}} , \lambda_{jf^{\rm SM}}  \right) - I_2 \left( \tau_{if^{\rm SM}} , \lambda_{jf^{\rm SM}}  
 \right) \biggr] \; ,
 \label{FfSMloopZg}
\eea
and
\bea
F_{ij} ( f^{\rm H} ) & = &  \frac{1}{16 \pi^2} \cdot N^c_{ f^{\rm H}} \cdot e Q_{ f^{\rm H}}  \cdot \frac{m_{ f^{\rm H}}} {v_\Phi} 
 \cdot C^{f^{\rm H}}_{Vj} \cdot \frac{-2}{m_{f^{\rm H}}} \cdot {{\mathcal O}^S_{2i} } \nonumber \\
 & \, & \times \biggl[ I_1 \left( \tau_{if^{\rm H}} , \lambda_{jf^{\rm H}}  \right) - I_2 \left( \tau_{if^{\rm H}} , \lambda_{jf^{\rm H}}  
 \right) \biggr] \; ,
 \label{FfHloopZg}
\eea
with $N^c_{f}$ being the color factor and $Q_f$ the electric charge of $f$ in unit of $e > 0$;
the vector couplings $C^f_{Vj}$ of quarks and leptons are listed in Table~\ref{tab:CCQV} 
and Table~\ref{tab:CCLV} respectively.

\begin{table}[htbp!]
\centering
\begin{tabular}{l|l}
\hline
$C^u_{Vj}$ & $ \frac{1}{2} \left[ \frac{g}{c_W} \left( \frac{1}{2} - \frac{4}{3} s^2_W \right)  {\mathcal O}^G_{1j} 
+ g_H  \left( + \frac{1}{2} \right) {\mathcal O}^G_{2j} +  {\frac{1}{2}}  g_X  {\mathcal O}^G_{3j} \right] $  \\
$C^d_{Vj}$ &  $ \frac{1}{2} \left[ \frac{g}{c_W} \left( - \frac{1}{2} + \frac{2}{3} s^2_W \right)  {\mathcal O}^G_{1j} 
+ g_H  \left( - \frac{1}{2} \right) {\mathcal O}^G_{2j} +    g_X \left( - {\frac{1}{2}} \right) {\mathcal O}^G_{3j} \right] $\\
$C^{u^H}_{Vj}$ &  $ \frac{1}{2} \left[ - \frac{g}{c_W} \left(  \frac{4}{3} \right) s^2_W {\mathcal O}^G_{1j}  + g_H  \left( - \frac{1}{2} \right) {\mathcal O}^G_{2j} +  {\frac{1}{2}}  g_X {\mathcal O}^G_{3j} \right] $ \\
$C^{d^H}_{Vj}$ &  $ \frac{1}{2} \left[  \frac{g}{c_W} \left( \frac{2}{3} \right) s^2_W {\mathcal O}^G_{1j}  + g_H  \left( + \frac{1}{2} \right) {\mathcal O}^G_{2j} +   g_X \left( - {\frac{1}{2}}  \right) {\mathcal O}^G_{3j} \right] $ \\
\hline
\end{tabular}
\caption{Coupling coefficients $C^f_{Vj}$ for quarks.}
\label{tab:CCQV}
\end{table}
\begin{table}[htbp!]
\centering
\begin{tabular}{l|l}
\hline
$C^\nu_{Vj}$ & $ \frac{1}{2} \left[ \frac{g}{c_W} \left( + \frac{1}{2} \right) {\mathcal O}^G_{1j} 
+ g_H  \left( + \frac{1}{2} \right) {\mathcal O}^G_{2j} +  {\frac{1}{2}}  g_X  {\mathcal O}^G_{3j} \right] $ \\
$C^e_{Vj}$ &   $  \frac{1}{2} \left[ \frac{g}{c_W} \left( - \frac{1}{2} + 2 s^2_W \right) {\mathcal O}^G_{1j}  
+ g_H  \left( - \frac{1}{2} \right) {\mathcal O}^G_{2j} +    g_X \left( - {\frac{1}{2}} \right) {\mathcal O}^G_{3j} \right] $ \\
$C^{\nu^H}_{Vj}$ &   $  \frac{1}{2} \left[ g_H  \left( - \frac{1}{2} \right) {\mathcal O}^G_{2j} + {\frac{1}{2}}   g_X  {\mathcal O}^G_{3j} \right] $ \\
$C^{e^H}_{Vj}$ &  $ \frac{1}{2} \left[ \frac{g}{c_W} \left( 2 \right) s^2_W {\mathcal O}^G_{1j}  + g_H  \left( + \frac{1}{2} \right) {\mathcal O}^G_{2j} +   g_X \left( -{\frac{1}{2}}  \right) {\mathcal O}^G_{3j}  \right] $ \\
\hline
\end{tabular}
\caption{Coupling coefficients $C^f_{Vj}$ for leptons.
}
\label{tab:CCLV}
\end{table}

There is only one charged Higgs $H^\pm$ in G2HDM. Thus $F^0_{ij} = F_{ij} ( H^\pm )$ with
\bea
F_{ij} (H^\pm) &=&  \frac{1}{16 \pi^2} \cdot e Q_{H^+} \cdot g_{h_i H^+H^-} \cdot g_{Z_j H^+ H^-} \cdot \frac{2}{m^2_{H^\pm}} \nonumber \\
&\times& 
I_1 \left( \tau_{i H^\pm} , \lambda_{j H^\pm}  \right) \; ,
\label{FCHloopZg}
\eea
where $Q_{H^+}  = +1$, and $g_{h_i H^+H^-}$ and $g_{Z_j H^+ H^-}$ are the $h_i H^+ H^-$ and $Z_j H^+ H^-$ couplings in the G2HDM 
respectively. Explicitly they are
\bea
\label{ghiCHCH}
g_{h_i H^+H^-} & = & \left( 2 \lambda_H - \lambda^\prime_H \right) v {\mathcal O}^S_{1i} 
+ \left( \lambda_{H \Phi} + \lambda^\prime_{H \Phi} \right) v_\Phi \mathcal{O}^S_{2i} \,,\;\;\; \;\; \;\\
g_{Z_j H^+H^-} & = &  \; \frac{1}{2} ( g \, c_W - g^\prime s_W ) {\mathcal O}^G_{1j} - \frac{1}{2} g_H {\mathcal O}^G_{2j}  + {\frac{1}{2} } g_X {\mathcal O}^G_{3j} \; .
\label{gZjCHCH}
\eea

\subsection{Decay Rates of $h_i \to \gamma \gamma$ 
and $h_i \to gg$}
\label{Ratehigammagamma}


The partial decay rate for $h_i \to \gamma  \gamma $ ($i=1,2$) is 
\be
\Gamma( h_i \to \gamma \gamma ) = \frac{1}{64 \pi} m_{h_i}^3 
\biggl\vert F^1_{ i }  + F^{1/2}_{ i }  + F^0_{ i }  \biggr\vert^2 \; ,
\label{Ratehiggs}
\ee
where
$F^1_{ i } = F_{ i }( W^\pm )$, 
$F^{1/2}_{ i } = \sum_{f^{\rm SM}} F_{ i }( f^{\rm SM} ) + 
\sum_{f^H} F_{ i }( f^H ) $, where $\sum_{f^{\rm SM}}$ and $\sum_{f^H}$ denote summation over all charged SM and heavy hidden fermions, and
$F^0_{ i } = F_{ i }( H^\pm )$ with
\begin{widetext}
\bea
\label{FWloopgg}
F_{ i } ( W^\pm ) & = & 
\frac{1}{16 \pi^2} \cdot e^2 \cdot g m_W \cdot \frac{-1}{m_W^2} \cdot {\mathcal O}^S_{1i} \nonumber \\
&& \times
\left[ 2 + 3 \tau_{iW} + 3 \tau_{iW} \left( 2 -  \tau_{iW} \right) f (  \tau_{iW} ) \right]
\,  ,  \\
\label{FfSMloopgg}
F_{ i } ( f^{\rm SM} ) & = & 
 \frac{1}{16 \pi^2} \cdot N^c_{ f^{\rm SM}} \cdot e^2 Q^2_{ f^{\rm SM}}  \cdot \frac{m_{ f^{\rm SM}}} {v} 
\cdot \frac{4}{m_{f^{\rm SM}}} \cdot {\mathcal O}^S_{1i} \nonumber \\
& & \times \left\{ \tau_{if^{\rm SM}} \left[ 1 + \left( 1 - \tau_{if^{\rm SM}} \right) f ( \tau_{if^{\rm SM}} ) \right] \right\}
\; , \\
\label{FfHloopgg}
F_{ i } ( f^H )  & = & 
 \frac{1}{16 \pi^2} \cdot N^c_{ f^{\rm H}} \cdot e^2 Q^2_{ f^{\rm H}}  \cdot \frac{m_{ f^{\rm H}}} {v_\Phi} 
\cdot \frac{4}{m_{f^{\rm H}}} \cdot {\mathcal O}^S_{2i} \nonumber \\
& & \times \left\{ \tau_{if^{\rm H}} \left[ 1 + \left( 1 - \tau_{if^{\rm H}} \right) f ( \tau_{if^{\rm H}} ) \right] \right\}
\; , \\
\label{FCHloopgg}
F_{ i } ( H^\pm ) & = & \frac{1}{16 \pi^2} \cdot e^2 \cdot g_{h_i H^+ H^-} \cdot \frac{-1 }{ m^2_{H^\pm} } \nonumber \\
&& \times
\left\{ \tau_{iH^\pm} \left[ 1 -  \tau_{iH^\pm} f( \tau_{iH^\pm}) \right] \right\}  \; .
\eea
\end{widetext}

The partial decay rate for $h_i \to gg $ $ (i=1,2)$ is
\be
\Gamma( h_i \to gg) = \frac{1}{8 \pi} m_{h_i}^3 
\biggl\vert G^{1/2}_{ i }   \biggr\vert^2 \; ,
\label{RatehiggsGlueGlue}
\ee
where $G^{1/2}_{ i } = \sum_{q^{\rm SM}} G_{ i }( q^{\rm SM} ) + 
\sum_{q^H} G_{ i }( q^H ) $ 
and
\bea
G_{ i }( q^{\rm SM}  ) & = &
 \frac{1}{16 \pi^2} \cdot \frac{ 1 }{2} \cdot g_s^2  \cdot \frac{m_{ q^{\rm SM}}} {v} 
\cdot \frac{4}{m_{q^{\rm SM}}} \cdot {\mathcal O}^S_{1i} \nonumber \\
& & \times \left\{ \tau_{iq^{\rm SM}} \left[ 1 + \left( 1 - \tau_{iq^{\rm SM}} \right) f ( \tau_{iq^{\rm SM}} ) \right] \right\}
\label{GqSMloopgg}\, ,
\eea
\bea
 G_{ i }( q^{\rm H} ) & = &
 \frac{1}{16 \pi^2} \cdot \frac{ 1 }{2} \cdot g_s^2  \cdot \frac{m_{ q^{\rm H}}} {v_\Phi} 
\cdot \frac{4}{m_{q^{\rm H}}} \cdot {\mathcal O}^S_{2i} \nonumber \\
& & \times \left\{ \tau_{iq^{\rm H}} \left[ 1 + \left( 1 - \tau_{iq^{\rm H}} \right) f ( \tau_{iq^{\rm H}} ) \right] \right\}
\label{GqHloopgg}\, .
\eea


\allowdisplaybreaks
\bibliographystyle{apsrev4-1}
\bibliography{main}

\end{document}